\documentclass[sigconf,nonacm]{aamas}

\usepackage{balance}
\usepackage{mathtools}
\usepackage{array}

\setcopyright{none}
\fancypagestyle{preprint}{
  \fancyhf{}
  \fancyfoot[C]{\thepage}

}

\title[On Symmetry for Characterizing Agent Heterogeneity in Multi-Agent Systems]
{On Symmetry for Characterizing Agent Heterogeneity in Multi-Agent Systems}

\author{Jianhong Wang$^{\dagger\ddagger}$, Jonathan Lawry$^\ddagger$}
\affiliation{
  \institution{$^{\dagger}$INFORMED-AI Hub, University of Bristol \\ $^{\ddagger}$School of Engineering Mathematics and Technology, University of Bristol
  }
  \city{Bristol}
  \country{United Kingdom}
  }
\email{\{jianhong.wang, J.Lawry\}@bristol.ac.uk}

\begin{abstract}
Understanding agent heterogeneity can inform the design of shared and agent-specific policy representations, however existing criteria provide no unified structural test. We study finite-horizon POSGs and define a type-compatible transformation as a policy-evaluation symmetry when complete policy evaluation commutes with it for every decentralized behavioral policy. Policy-evaluation symmetries form a group for which agent orbits define structural agent types. This classification does not depend on the particular policy chosen or the distribution of initial states. We prove that every componentwise environment automorphism is a policy-evaluation symmetry, but that the converse does not hold. Applying the framework to cooperative Dec-POMDPs shows that agents can belong to different structural types even when they share a common reward. For prescribed weighted reward aggregation, we derive an exact value identity and sufficient conditions for the resulting Dec-POMDP to inherit the policy-evaluation symmetries of the source POSG. Finally, we give sufficient conditions under which repeated model factorizations induce policy-evaluation symmetry, and show that preserving an optimal Boltzmann policy set neither implies nor follows from this symmetry in general.
\end{abstract}

\keywords{multi-agent systems, agent heterogeneity, symmetry, POSGs, Dec-POMDPs}

\newcommand{\RAdm}{\mathcal R_{\mathrm{adm}}(\mathcal M)}
\newcommand{\Geval}{G_{\mathcal M}^{\mathrm{eval}}}
\newcommand{\Gevali}{G_{\mathcal M}^{\mathrm{eval},\rho_0}}
\newcommand{\AutEnv}{\operatorname{Aut}_{\mathrm{env}}}
\newcommand{\Peval}{\mathcal P_{\mathrm{eval}}}

\begin{document}
\maketitle
\pagestyle{preprint}
\thispagestyle{preprint}

\section{Introduction}
\label{sec:introduction}

Multi-agent systems can describe interacting decision-makers in real-world
problems, such as cooperative robot
control~\citep{bettini2023heterogeneous} and voltage regulation in power
distribution networks~\citep{wang2021voltage}, where agents' capabilities,
information, objectives, and behavior may differ.  As a result, characterizing these
differences is necessary, which can inform the design of policy to make decisions.

Symmetry identifies transformations that preserve relevant system
properties, a principle used across physical and computational sciences.
In power-grid swing equations, exchanging generator labels preserves the
dynamics when the corresponding generator parameters and network couplings
are unchanged~\citep{molnar2021asymmetry}.  In molecular
modeling, energy is invariant to rigid rotations, translations, and atom
reindexing, properties incorporated into SchNet~\citep{schutt2017schnet}.
These examples inspires a possibility to identify corresponding structural regularities
in multi-agent systems by symmetries.

Multi-agent reinforcement learning (MARL), as a learning framework for multi-agent systems, already exploits symmetry.
Distributed policy networks that encode global symmetries empirically
improve data efficiency on symmetric cooperative
tasks~\citep{vanderpol2022multiagent}.  Other-play uses environment symmetries
to reduce dependence on arbitrary conventions in zero-shot
coordination~\citep{hu2020otherplay}, while expected return symmetries extend
the symmetry class used for this purpose~\citep{muglich2025expected}.
However, their focus is task-specific (improving learning and coordination), leaving the connection
to structural agent heterogeneity an open question.

Broader formal accounts describe component-preserving model
automorphisms~\citep{kang2012symmetries,treutlein2021formalism}.
Braun et al.~\citep{BraunGL022} use counting and isomorphic symmetries for
compact partitioned Dec-POMDPs, with agent-number tractability relying on
the stronger isomorphic factorization.  Complementary work classifies
physical and behavioral heterogeneity in multi-robot
learning~\citep{bettini2023heterogeneous}, measures behavioral differences
through system neural diversity~\citep{bettini2025system}, and distinguishes
observation, transition, objective, and policy
heterogeneity~\citep{hu2026heterogeneity}.  These perspectives identify
particular model regularities or dimensions of difference, however, they are still with a gap to the chracteristics of multi-agent systems.

We therefore ask: \emph{can we use symmetries to determine structural agent
heterogeneity?}  Specifically, when can agents be exchanged through
compatible transformations without changing how every decentralized policy is
evaluated?

To answer the above research questions, we ground our theoretical framework in finite-horizon partially observable stochastic games (POSGs).
A transformation is a \emph{policy-evaluation symmetry} if evaluating any
transformed policy gives the correspondingly transformed value profile, at every
state--joint-history pair.  These symmetries form a group whose agent orbits
define \emph{structural agent types}: one orbit means structural homogeneity,
and multiple orbits mean structural heterogeneity.  For the specified game
and admissible transformations, this classification is independent of the selected
policy and initial-state distribution.

Our contributions are as follows: (1) We establish the policy-evaluation symmetry group and its structural
agent types, allowing compatible agent, state, action, and observation transformations.  We distinguish policy-evaluation symmetry, which concerns every decentralized policy, from the equivariance of a selected policy. (2) We prove that preserving the transition, observation, and reward functions componentwise guarantees policy-evaluation symmetry, but the reverse is not necessary.  A finite two-stage POSG yields a strictly finer componentwise agent partition to show the necessity of the policy-evaluation definition. (3) We show that Dec-POMDPs can remain structurally heterogeneous, although with a common reward.  For a derived Dec-POMDP with a prescribed weighted common reward, we derive an exact weighted-value identity and sufficient conditions for inheriting the source game's symmetries. We further compare the framework with existing heterogeneity categories,
counting and isomorphic symmetry, and expected return symmetry, making the
assumptions and limitations of each comparison explicit.

\section{Related Work}
\label{sec:related-work}

\subsection{Taxonomies of Agent Heterogeneity}
\label{sec:related-heterogeneity}

Bettini et al.~\cite{bettini2023heterogeneous} distinguish physical from
behavioral heterogeneity in multi-robot and multi-agent systems.  They further
separate behavioral heterogeneity according to whether agents share an
objective, yielding five physical--behavioral subclasses.  This taxonomy is a
high-level classification rather than a measure of degree.  Their later system
neural diversity measure addresses the latter question for behavior by
averaging pairwise behavioral distances~\cite{bettini2025system}.  Hu et
al.~\cite{hu2026heterogeneity} give a finer agent-level MARL taxonomy based on
observation, response-transition, effect-transition, objective, and policy
heterogeneity.

Table~\ref{tab:heterogeneity-comparison} compares these prior categories and
the system neural diversity measure with
the closest concepts in our framework.  For physical, observation, transition,
and objective heterogeneity, the associations are conditional on the relevant
agent property being represented in the POSG.  Behavioral and policy
heterogeneity are instead compared through equivariance of a selected policy
relative to chosen policy-evaluation symmetries.  Bettini et al.'s five
subclasses combine physical and behavioral heterogeneity, with behavioral
heterogeneity further distinguished by whether agents share an objective.
The table therefore lists these underlying distinctions rather than repeating
all five combinations.

\begin{table}[t]
\caption{Relation of prior heterogeneity categories and measures to our
symmetry-based concepts.  Note that these associations are not claims of formal
equivalence.}
\label{tab:heterogeneity-comparison}
\centering
\footnotesize
\setlength{\tabcolsep}{3pt}
\renewcommand{\arraystretch}{1.13}
\begin{tabular}{@{}
  >{\raggedright\arraybackslash}p{.34\columnwidth}
  >{\raggedright\arraybackslash}p{.62\columnwidth}@{}}
\toprule
\textbf{Prior category or measure} & \textbf{Relation to our framework} \\
\midrule
\textbf{Physical heterogeneity}~\citep{bettini2023heterogeneous}
& Physical differences can be tested through a \textbf{componentwise
environment automorphism} when their consequences are encoded in the POSG. \\\\
\textbf{Observation heterogeneity}~\citep{hu2026heterogeneity}
& \textbf{Observation-preservation condition} within a
componentwise environment automorphism. \\\\
\textbf{Response- and effect-transition heterogeneity}~\citep{hu2026heterogeneity}
& \textbf{Transition-preservation condition} within a
componentwise environment automorphism, provided that the required local
transition factorization is specified. \\\\
\textbf{Objective heterogeneity}~\citep{bettini2023heterogeneous,hu2026heterogeneity}
& \textbf{Reward-preservation condition} within a
componentwise environment automorphism. \\\\
\textbf{Behavioral or policy heterogeneity}~\citep{bettini2023heterogeneous,hu2026heterogeneity}
& \textbf{Equivariance of a selected policy} characterizes behavioral
homogeneity among agents related by the chosen policy-evaluation symmetries, while failure relative to a specified symmetry expresses policy-relative behavioral
differentiation. \\\\
\textbf{System neural diversity}~\citep{bettini2025system}
& \textbf{Graded comparison of selected policies} that complements
the exact classification into structural agent types. \\
\bottomrule
\end{tabular}
\end{table}

These associations locate earlier categories within the symmetry framework, but they do not identify those categories with structural agent types.  A
componentwise environment automorphism that exchanges two agents guarantees
that they have the same structural agent type.  However, failure of a
componentwise preservation equation for a candidate transformation neither rules
out that transformation as a policy-evaluation symmetry nor establishes that the
agents have different structural agent types.  Likewise, different behavior
under one selected policy does not by itself establish different structural
agent types.  Graded behavioral measures such as system neural
diversity~\citep{bettini2025system} are complementary to the structural
agent-type partition.  Appendix~\ref{app:heterogeneity-comparison} explains
the assumptions and direction of each association in
Table~\ref{tab:heterogeneity-comparison}.

\subsection{Symmetry in Multi-Agent Systems}
\label{sec:related-symmetry}

Symmetry in normal-form games is commonly expressed through permutations of
players and strategies that preserve payoff profiles.  Cao and
Yang~\cite{cao2018symmetric} formalize player and strategy symmetry groups and
distinguish ordinary, renaming, and name-irrelevant symmetric games.  Koriche
et al.~\cite{koriche2017constraint} distinguish structural symmetry from
minimax symmetry in finite-horizon stochastic games.  Their semantic notion
preserves an optimal minimax value, whereas ours evaluates every decentralized
policy in a partially observable game.  Tewolde et
al.~\cite{tewolde2025computing} study the complexity of detecting player and
action symmetries in normal-form games, deriving graph-automorphism and
graph-isomorphism completeness results, and analyze Nash equilibria that
respect given symmetries.  These traditions motivate treating agent
interchangeability as an orbit question but do not directly address our
all-policy setting with type-compatible state, observation, and history transformations.

For sequential models, Kang and Kim~\cite{kang2012symmetries} study
symmetries of POMDPs and POSGs through graph automorphisms that preserve the
model components.  In cooperative settings, Hu et
al.~\cite{hu2020otherplay} call component-preserving transformations
\emph{equivalence mappings} and use them to randomize conventions in
other-play.  Treutlein et al.~\cite{treutlein2021formalism} formulate
Dec-POMDP isomorphisms and automorphisms using bijections of agents, states,
actions, and observations that preserve transitions, observations, rewards,
and the initial distribution.  These definitions use different model
conventions and admissible transformations.  After translating them into our
setting, their requirements to preserve the transition, observation, and
reward functions become instances of the componentwise environment
automorphisms studied in Section~\ref{sec:componentwise}.  By
Theorem~\ref{thm:componentwise-certificate}, these conditions are sufficient,
but not necessary, for policy-evaluation symmetry.  We instead define
policy-evaluation symmetry directly by
\(
\mathcal E_{\mathcal M}(g\cdot\pi)
=g\cdot\mathcal E_{\mathcal M}(\pi)
\)
for every decentralized policy \(\pi\): transforming a policy before
evaluation must give the same value profile as evaluating it first and then
transforming its values.

Other work captures different consequences or representations of symmetry.
At the policy-architecture level, van der Pol et
al.~\cite{vanderpol2022multiagent} factor global cooperative-system
symmetries into local transformations and agent permutations to construct
distributed equivariant policy networks.  This exploits a supplied symmetry
in policy learning, whereas our structural agent types are defined before selecting
a policy or architecture.  At the representation level,
Braun et al.~\cite{BraunGL022} use counting and isomorphic symmetry to obtain
compact lifted Dec-POMDP representations.  Muglich et
al.~\cite{muglich2025expected} introduce expected return symmetries, which
preserve self-play-optimal Boltzmann-exploratory policies at a fixed
temperature for zero-shot coordination.  Our structural agent types instead use the orbits
of a solution-independent policy-evaluation symmetry group that holds for every decentralized policy.

\section{Model and Type-Compatible Transformations}
\label{sec:setting}

We consider a finite-horizon partially observable stochastic game
(POSG)~\citep{hansen2004dynamic}, as follows:
\[
\mathcal M=\bigl(N,\mathcal S,(\mathcal A_i)_i,(\mathcal O_i)_i,
P,Z,(r_i)_i,\rho_0,T,\gamma\bigr),
\]
where \(N=\{1,\ldots,n\}\) is the finite nonempty agent set, and
\(\mathcal S\), \(\mathcal A_i\), and \(\mathcal O_i\) are finite and
nonempty for every \(i\in N\). Moreover, the finite horizon is \(T<\infty\) and
the discount factor is \(\gamma\in(0,1]\). The transition function is
\(P(s'\mid s,a)\), the joint observation function is
\(Z(o'\mid s',a)\), and \(r_i(s,a,s')\) is agent \(i\)'s reward function.
At the initial time step \(t=0\), the state is drawn from the initial
distribution \(\rho_0\), so \(s_0\sim\rho_0\), and every agent has the empty
history \(h_i^0=()\).  For \(t\geq1\),
the local history of agent \(i\) is defined as:
\[
h_i^t=(a_i^0,o_i^1,\ldots,a_i^{t-1},o_i^t)
\in\mathcal H_i^t=(\mathcal A_i\times\mathcal O_i)^t.
\]

The policy domain contains all decentralized behavioral product policies, such that:
\[
\Pi_{\mathrm{dec}}
=\prod_{i\in N}\prod_{t=0}^{T-1}
  \prod_{h_i^t\in\mathcal H_i^t}\Delta(\mathcal A_i),
\qquad
\pi^t(a\mid h^t)=\prod_i\pi_i^t(a_i\mid h_i^t).
\]
Agents may randomize independently at their local histories, but we exclude a
common random signal, such as a shared seed or public correlation device
\citep{aumann1974subjectivity,bernstein2009policy}.\footnote{Conditioning local policies on the same random draw can produce non-product joint action laws that the factorization above cannot represent, thereby enlarging the admissible policy domain. Because policy-evaluation symmetry quantifies over this domain, admitting correlated policies can change the symmetry group and its induced structural agent types.}

Let \(\mathcal X_t=\mathcal S\times\prod_i\mathcal H_i^t\).  Complete
policy evaluation assigns a value to every valid state--joint-history pair,
including pairs unreachable under a selected policy.  With
\(V_{i,T}^{\pi}\equiv0\), for \(t<T\) it is defined by the following Bellman recursion:
\begin{align}
V_{i,t}^{\pi}(s,h^t)
=\sum_{a,s',o'}&\pi^t(a\mid h^t)P(s'\mid s,a)Z(o'\mid s',a)
\notag\\[-2pt]
&\quad\cdot\bigl[r_i(s,a,s')+\gamma
V_{i,t+1}^{\pi}(s',h^t,a,o')\bigr].
\label{eq:bellman}
\end{align}
The final transition reward is therefore included at \(t=T-1\).  Write
\[
\mathcal E_{\mathcal M}:\Pi_{\mathrm{dec}}\to\mathcal V,
\qquad
\mathcal E_{\mathcal M}(\pi)=V^\pi,
\]
for the operator returning all agents' values at all time steps and pairs, where
\(\mathcal V\) is the space of all function profiles
\(V=(V_{i,t})\), where \(V_{i,t}:\mathcal X_t\to\mathbb R\) for
\(i\in N\) and \(0\leq t\leq T\), with \(V_{i,T}\equiv0\).
Each policy \(\pi\) determines a particular profile \(V^\pi\in\mathcal V\)
through Equation~\eqref{eq:bellman}.  The expected return under the initial
distribution is as follows:
\begin{equation}
\label{eq:expected-return}
J_i(\pi)
=\mathbb E_{s_0\sim\rho_0}
\bigl[V_{i,0}^\pi(s_0,()^N)\bigr],
\end{equation}
where \(()^N\) is the empty joint history.

\paragraph{Type-compatible transformations.}
A type-compatible transformation is a complete tuple
\[
g=\bigl(\sigma_g,\phi_g^{\mathcal S},
(\phi_{g,i}^{\mathcal A})_i,(\phi_{g,i}^{\mathcal O})_i\bigr),
\]
where \(\sigma_g\) permutes agents, \(\phi_g^{\mathcal S}\) bijects the
state space, and
\(\phi_{g,i}^{\mathcal A}:\mathcal A_i\to
\mathcal A_{\sigma_g(i)}\) and
\(\phi_{g,i}^{\mathcal O}:\mathcal O_i\to
\mathcal O_{\sigma_g(i)}\) are bijections.  It acts on joint actions by
\([\smash{g\cdot a}]_{\sigma_g(i)}=\phi_{g,i}^{\mathcal A}(a_i)\), and
analogously on joint observations.  For \(t\geq1\), the induced local-history
map \(\Phi_{g,i}^{\mathcal H,t}:\mathcal H_i^t\to
\mathcal H_{\sigma_g(i)}^t\) is expressed as follows:
\[
\begin{aligned}
&\Phi_{g,i}^{\mathcal H,t}
(a_i^0,o_i^1,\ldots,a_i^{t-1},o_i^t)\\
&\quad=\bigl(
\phi_{g,i}^{\mathcal A}(a_i^0),
\phi_{g,i}^{\mathcal O}(o_i^1),\ldots,
\phi_{g,i}^{\mathcal A}(a_i^{t-1}),
\phi_{g,i}^{\mathcal O}(o_i^t)
\bigr),
\end{aligned}
\]
and \(\Phi_{g,i}^{\mathcal H,0}(())=()\).  For a joint history
\(h^t=(h_i^t)_{i\in N}\), define
\[
[g\cdot h^t]_{\sigma_g(i)}
=\Phi_{g,i}^{\mathcal H,t}(h_i^t).
\]
Thus agent \(i\)'s transformed local history occupies coordinate
\(\sigma_g(i)\) in the joint history.
The induced state--joint-history map is
\[
\tau_g^t:\mathcal X_t\to\mathcal X_t,
\qquad
\tau_g^t(s,h^t)=\bigl(\phi_g^{\mathcal S}(s),g\cdot h^t\bigr).
\]
Thus \(\tau_g^t\) transforms the state and joint history.  It preserves the time index\footnote{The time sequence is not changed, e.g., a time sequence \((t_1, t_2, t_3)\) would still be \((t_1, t_2, t_3)\) after the transformation.} and does not perform a
state transition to generate the next state.  In particular, for the empty joint history \(()^N\),
\(\tau_g^0(s,()^N)=(\phi_g^{\mathcal S}(s),()^N)\).

We fix an application-specified subgroup \(\RAdm\) of complete
type-compatible transformations arising from representation structure
fixed for the application. Examples include permutations of an
entity-factored multi-agent representation~\citep{hao2023boosting}, node
relabelings of a graph-structured representation~\citep{keriven2019universal},
or geometric state--action transformations such as rotations and
reflections~\citep{vanderpol2020mdp}. Membership in \(\RAdm\) supplies a candidate
transformation; an environment or policy-evaluation symmetry must also
satisfy the corresponding preservation conditions.

The induced transformations of policies and values are specified pointwise
by the following equations:
\begin{align}
(g\cdot\pi)_{\sigma_g(i)}^t
\bigl(\phi_{g,i}^{\mathcal A}(a_i)\mid
\Phi_{g,i}^{\mathcal H,t}(h_i^t)\bigr)
&=\pi_i^t(a_i\mid h_i^t),
\label{eq:policy-action}\\
(g\cdot V)_{\sigma_g(i),t}(\tau_g^t x_t)
&=V_{i,t}(x_t).
\label{eq:value-action}
\end{align}
These transformations define left group actions.  At each fixed joint
history, the transformed joint action distribution equals the product of the
transformed local distributions.  Because the induced policy transformation
only relabels and reindexes the local policies, \(g\cdot\pi\) remains a
decentralized product policy in \(\Pi_{\mathrm{dec}}\).
Appendix~\ref{app:foundations} proves these action properties together with
the subgroup and orbit facts used in the following sections.

\section{Policy-Evaluation Symmetry and Structural Agent Types}
\label{sec:policy-evaluation}

\begin{definition}[Policy-evaluation symmetry]
\label{def:policy-evaluation}
An admissible transformation \(g\in\RAdm\) is a \emph{policy-evaluation
symmetry} if
\[
\mathcal E_{\mathcal M}(g\cdot\pi)
=g\cdot\mathcal E_{\mathcal M}(\pi)
\qquad\text{for every }\pi\in\Pi_{\mathrm{dec}}.
\]
Equivalently,
\[
V_{\sigma_g(i),t}^{g\cdot\pi}(\tau_g^t x_t)
=V_{i,t}^{\pi}(x_t)
\]
for every policy, agent, time step, and \(x_t\in\mathcal X_t\).  The set of all
such transformations is denoted \(\Geval\).
\end{definition}

For a normal-form game embedded as a one-stage POSG, this definition reduces
to payoff preservation under admissible player and action relabelings
(see Proposition~\ref{prop:supp-normal-form}).

The requirement that the identity hold for every policy \(\pi\in\Pi_{\mathrm{dec}}\) is essential.  The definition says that
transforming any policy and then evaluating it gives the correspondingly
transformed complete value profile.  It does not require \(V^\pi\) to be
invariant under \(g\), because in general the comparison is between \(\pi\)
and \(g\cdot\pi\).
The group-action laws imply that \(\Geval\leq\RAdm\).

\begin{definition}[Structural agent type]
\label{def:structural-type}
Agents \(i,j\in N\) have the same \emph{structural agent type} when
\[
i\sim_{\mathrm{eval}} j
\quad\Longleftrightarrow\quad
\text{some }g\in\Geval\text{ satisfies }\sigma_g(i)=j.
\]
The structural partition is the orbit partition
\[
\Peval(\mathcal M)
=\mathcal P(\Geval)
=\bigl\{\{\sigma_g(i):g\in\Geval\}:i\in N\bigr\}.
\]
\end{definition}

Writing \(\Peval(\mathcal M)=\{T_1,\ldots,T_m\}\), each \(T_k\) is one
structural agent type.  The multi-agent system is structurally homogeneous if \(m=1\) and
structurally heterogeneous if \(m>1\). Heterogeneity therefore describes the
complete partition, rather than one particular type.

Because \(\Geval\) acts on \(N\) through
the agent maps \(i\mapsto\sigma_g(i)\), \(\sim_{\mathrm{eval}}\) is the
standard orbit equivalence relation~\citep[Chap.~14.1]{judson1994abstract}.\footnote{Specifically, \(e\in\Geval\) satisfies \(\sigma_e(i)=i\), giving reflexivity.
If \(g\in\Geval\) maps \(i\) to \(j\), then \(g^{-1}\in\Geval\) maps \(j\)
to \(i\), giving symmetry.  If \(g,k\in\Geval\) map \(i\) to \(j\) and
\(j\) to \(\ell\), respectively, then \(kg\in\Geval\) and
\(\sigma_{kg}(i)=\sigma_k(\sigma_g(i))=\ell\), giving transitivity.} Mathematically, the case \(m=1\) means that the action of \(\Geval\) on \(N\)
is transitive: for every pair of agents \(i,j\in N\), some \(g\in\Geval\)
satisfies \(\sigma_g(i)=j\).

To obtain an equivariance relation for expected returns \(J_i(\pi)\) under the initial distribution, we additionally require the state transformation to preserve \(\rho_0\). We therefore define the subgroup, as follows:
\[
\Gevali
=\left\{g\in\Geval:
\rho_0\bigl(\phi_g^{\mathcal S}(s)\bigr)=\rho_0(s)
\text{ for every }s\in\mathcal S\right\},
\]
where the state transformation must preserve the initial distribution.  Therefore, every
\(g\in\Gevali\) satisfies the following equation:
\[
J_{\sigma_g(i)}(g\cdot\pi)=J_i(\pi).
\]

\paragraph{Policy equivariance is downstream.}
For an arbitrary subgroup \(H\leq\Geval\), we call a selected joint policy \(H\)-equivariant when
\(g\cdot\pi=\pi\), for every \(g\in H\).  Pointwise, this is represented by the following equation:
\[
\pi_{\sigma_g(i)}^t
\bigl(\phi_{g,i}^{\mathcal A}(a_i)\mid
\Phi_{g,i}^{\mathcal H,t}(h_i^t)\bigr)
=\pi_i^t(a_i\mid h_i^t).
\]
For every \(g\in H\), policy-evaluation symmetry gives
\(g\cdot V^\pi=V^{g\cdot\pi}\).  If the selected policy is
\(H\)-equivariant, then \(g\cdot\pi=\pi\), so its value profile is also
unchanged:
\[
g\cdot V^\pi=V^{g\cdot\pi}=V^\pi.
\]The reverse implication need not hold. The equality \(g\cdot V^\pi=V^\pi\) for one selected policy does not imply that \(g\cdot\pi=\pi\), nor does it establish the following identity for every policy:
\[
V^{g\cdot\eta}=g\cdot V^\eta
\quad\text{for every }\eta\in\Pi_{\mathrm{dec}},
\]as required for \(g\) to be a policy-evaluation symmetry.

\section{Componentwise Environment Automorphisms}
\label{sec:componentwise}

Componentwise primitive-preserving transformations,\footnote{Agent-index conventions differ: the POSG automorphisms of \citet{kang2012symmetries} and Dec-POMDP automorphisms of \citet{treutlein2021formalism} include an agent bijection; the equivalence mappings of \citet{hu2020otherplay} and transformation group of \citet{muglich2025expected} fix agent indices. Our type-compatible transformations permit agent permutations, following the former convention.} such as transformations that separately preserve the transition, observation, and reward functions, appear as POSG automorphisms or
symmetries~\citep{kang2012symmetries}, Dec-POMDP equivalence
mappings~\citep{hu2020otherplay}, Dec-POMDP
automorphisms~\citep{treutlein2021formalism}, and environment
symmetries~\citep{muglich2025expected}.  We use the name
\emph{componentwise environment automorphism} for this notion in POSGs,
emphasizing its contrast with policy-evaluation symmetry.  Unlike the
Dec-POMDP automorphisms of \citet{treutlein2021formalism}, preservation of the
initial distribution is imposed separately in our framework, as we introduced in Section~\ref{sec:policy-evaluation}.

\begin{definition}[Componentwise environment automorphism~\citep{kang2012symmetries}]
\label{def:componentwise}
An admissible transformation \(g\) is a componentwise environment
automorphism when, for every well-typed \(s,s',a,o'\), and \(i\),
\begin{align}
P(\phi_g^{\mathcal S}(s')\mid\phi_g^{\mathcal S}(s),g\cdot a)
&=P(s'\mid s,a),
\label{eq:component-P}\\
Z(g\cdot o'\mid\phi_g^{\mathcal S}(s'),g\cdot a)
&=Z(o'\mid s',a),
\label{eq:component-Z}\\
r_{\sigma_g(i)}(\phi_g^{\mathcal S}(s),g\cdot a,
\phi_g^{\mathcal S}(s'))
&=r_i(s,a,s').
\label{eq:component-r}
\end{align}
Write \(\AutEnv(\mathcal M)\) for the set of componentwise environment
automorphisms.
\end{definition}

\begin{theorem}[Componentwise environment automorphisms are sufficient]
\label{thm:componentwise-certificate}
The componentwise environment automorphisms form a subgroup and
\[
\AutEnv(\mathcal M)\leq\Geval,
\qquad
\mathcal P(\AutEnv(\mathcal M))
\preceq\Peval(\mathcal M),
\]
where \(\preceq\) means finer than or equal to.
\end{theorem}

\paragraph{Proof sketch.}
The identity transformation preserves all three environment components.  If two
transformations preserve them, applying the preservation equations successively
shows that their composition also preserves them.  Substituting inverse
images into the same equations shows that each inverse preserves them.
Hence \(\AutEnv(\mathcal M)\) is a subgroup.

It remains to prove the inclusion in \(\Geval\).  At \(t=T\), both values
in Definition~\ref{def:policy-evaluation} are zero.  Assume their equality at
time step \(t+1\), and expand the transformed Bellman recursion at time step \(t\).
Make the bijective substitutions
\[
b=g\cdot a,
\qquad
\bar s'=\phi_g^{\mathcal S}(s'),
\qquad
\bar o'=g\cdot o'.
\]
The policy-transformation rule in Equation~\eqref{eq:policy-action} matches
the transformed joint-action probability with the source probability.
Equations~\eqref{eq:component-P}--\eqref{eq:component-r} replace the
transition, observation, and reward terms with their source counterparts.
The history transformation identifies corresponding next state--joint-history
pairs.  The induction hypothesis then equates their values at time step
\(t+1\) under the transformed and source policies.
The transformed Bellman sum therefore
reduces to Equation~\eqref{eq:bellman}.  Consequently,
\(
V_{\sigma_g(i),t}^{g\cdot\pi}(\tau_g^t x_t)
=V_{i,t}^{\pi}(x_t)
\)
for every \(g\in\AutEnv(\mathcal M)\), policy \(\pi\), agent \(i\), time step
\(t\), and \(x_t\in\mathcal X_t\).  Thus, every componentwise environment
automorphism is a policy-evaluation symmetry, proving
\(\AutEnv(\mathcal M)\leq\Geval\).
The subgroup inclusion also gives the partition refinement, as follows:
\[
\mathcal P(\AutEnv(\mathcal M))
\preceq\Peval(\mathcal M),
\]
because every orbit under \(\AutEnv(\mathcal M)\) is contained in an orbit
under \(\Geval\).
Appendix~\ref{app:componentwise-proofs} gives the complete proof of
Theorem~\ref{thm:componentwise-certificate}.

The inclusion can yield a strict refinement, as follows:

\begin{proposition}[Strictness of the componentwise sufficient condition]
\label{prop:strict-witness}
There exists a finite, full-support, two-agent, two-stage POSG for
which
\[
\AutEnv(\mathcal M)\lneq\Geval
\quad\text{and}\quad
\mathcal P(\AutEnv(\mathcal M))
\prec\Peval(\mathcal M).
\]
Here \(\prec\) means strictly finer than.
\end{proposition}

\paragraph{Proof sketch.}
Appendix~\ref{app:strict-witness} constructs a full-support two-stage POSG in
which exchanging the agents changes the transition distribution associated
with some joint actions.  The exchange is therefore not a componentwise
environment automorphism.  Nevertheless, the affected transition
distributions produce the same expected successor-state reward and
marginalized next-observation distribution.  The Bellman recursion preserves
complete policy evaluation under the exchange for every decentralized policy.
The reward construction also rules out every alternative componentwise
transformation that exchanges the agents.  Hence, the componentwise partition
separates the agents, whereas the policy-evaluation partition places them in
one structural agent type.

\begin{figure*}[t]
\centering
\includegraphics[width=\textwidth]{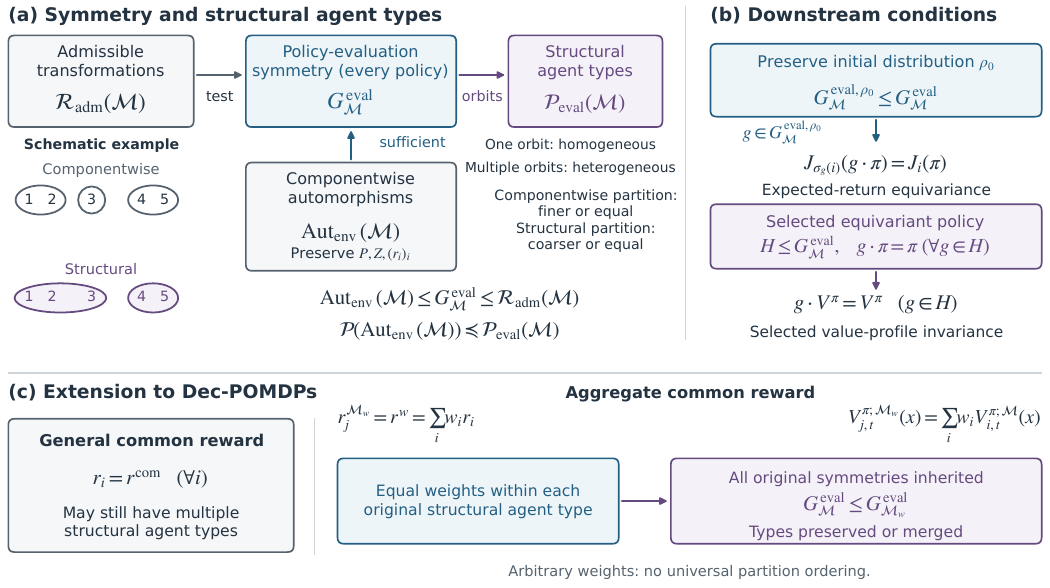}
\caption{Hierarchy of key concepts and results. $\leq$ denotes
subgroup inclusion. $\preceq$ means finer than or equal to.
The schematic inset illustrates orbit refinement: each componentwise orbit
lies within one structural agent type. Initial-distribution preservation and
selected-policy equivariance impose separate conditions for expected-return
equivariance and invariance of a selected value profile.
Panel (c) shows that a common reward need not imply structural homogeneity.
For the derived common-reward model, equal weights within each original
structural type guarantee symmetry inheritance, so types are preserved or
merged.}
\Description{Componentwise environment automorphisms are a subgroup of
policy-evaluation symmetries, which are a subgroup of admissible
transformations. Agent orbits under policy-evaluation symmetry define structural
types. A schematic example shows three componentwise parts and two
structural types, with non-singleton parts in both partitions. The
componentwise partition can be finer. Separate conditions on the
initial distribution and selected policy yield expected-return equivariance
and selected value-profile invariance, respectively.
A full-width third panel summarizes the cooperative extensions.
A common-reward model may still have several structural agent types.
Weighted reward aggregation gives a weighted-value identity. Equal weights
within each original structural type preserve all original policy-evaluation
symmetries, so structural types are preserved or merged. Without this
weight condition, no universal partition ordering is guaranteed.}
\label{fig:symmetry-hierarchy}
\end{figure*}

\section{Extension to Dec-POMDPs}
\label{sec:cooperative}

We apply the policy-evaluation symmetry and orbit framework to two settings of Dec-POMDP~\citep{bernstein2002complexity}.  The first is a Dec-POMDP with a general common reward.  The second is a derived Dec-POMDP formed by aggregating the individual rewards of a POSG.

\subsection{General Common Rewards}

A finite-horizon POSG specializes to a Dec-POMDP when all agents share one
reward function, \(r_i=r^{\mathrm{com}}\) for every \(i\in N\).  The common
terminal condition and Bellman recursion then imply that
\[
V_{i,t}^{\pi}(x)=V_{j,t}^{\pi}(x)
\qquad
\text{for every }i,j,\pi,t,\text{ and }x\in\mathcal X_t.
\]
This equality compares the agents' value coordinates under the same policy
and at the same state--joint-history input. However, it does not show that their
decision-making roles are interchangeable.  By
Definition~\ref{def:structural-type}, agents \(i\) and \(j\) have the same
structural agent type only if some \(g\in G_{\mathcal M}^{\mathrm{eval}}\)
maps \(i\) to \(j\).  Such a transformation must include compatible
bijections between the agents' action and observation spaces and must preserve
complete policy evaluation for every decentralized policy after exchanging
the agents.  A common reward alone guarantees neither these bijections nor
symmetry of the transition and observation distributions.  For example,
in a one-stage binary-action model with the common reward
\(r(a_1,a_2)=a_1\), both agents receive the same payoff, but exchanging them
changes which agent's policy determines that payoff. The detailed proof of the above statement is given in Proposition~\ref{prop:supp-common-reward} in Appendix~\ref{app:cooperative-proofs}.

Hence, a Dec-POMDP may still have multiple structural agent types despite the
common reward, and it is structurally
homogeneous exactly when \(G_{\mathcal M}^{\mathrm{eval}}\) acts transitively
on the agent set \(N\). 

\subsection{Aggregate Common Rewards}

Let \(\mathcal M\) be a POSG with agent-specific reward functions
\((r_i)_{i\in N}\).  Fix nonnegative weights
\(w\in\mathbb R_{\geq0}^N\setminus\{0\}\) and define the aggregate common reward, as follows:
\[
r^w(s,a,s')=\sum_{i\in N}w_i r_i(s,a,s').
\]
Here \(r^w\) is only a scalar function constructed from the rewards of
\(\mathcal M\), so defining it does not replace the source reward functions
in \(\mathcal M\).  We construct another Dec-POMDP
\(\mathcal M_w\) by retaining the agents, state, action, and observation
spaces, transition and observation distributions, initial distribution,
horizon, discount factor, and policy domain of \(\mathcal M\), while assigning
\(r^w\) to every agent:
\[
r_j^{\mathcal M_w}=r^w
\qquad\text{for every }j\in N.
\]
Thus, \(\mathcal M\) and \(\mathcal M_w\) differ only in their reward
specifications.  We test both models using the same declared admissible
transformations, so any difference between their symmetry groups comes from
replacing the individual rewards with the common reward \(r^w\), rather than
from allowing different candidate transformations.

\begin{theorem}[Weighted aggregation and inherited symmetry]
\label{thm:weighted-aggregation}
For every agent \(j\), policy \(\pi\), time step \(t\), and \(x\in\mathcal X_t\),
\[
V_{j,t}^{\pi;\mathcal M_w}(x)
=\sum_{i\in N}w_iV_{i,t}^{\pi;\mathcal M}(x).
\]
Moreover,
\[
\bigl\{g\in G_{\mathcal M}^{\mathrm{eval}}:
w_{\sigma_g(i)}=w_i\ \text{for all }i\bigr\}
\leq G_{\mathcal M_w}^{\mathrm{eval}}.
\]
At the componentwise level,
\[
\left\{g\in\AutEnv(\mathcal M):
w_{\sigma_g(i)}=w_i\ \text{for all }i\right\}
\leq\AutEnv(\mathcal M_w).
\]
\end{theorem}

\paragraph{Proof sketch.}
At \(t=T\), both sides of the value identity are zero.  Suppose the identity
holds at \(t+1\).  For \(x=(s,h^t)\), write \(x'\) for the successor
state--joint-history input determined by \(s',a,o'\).  Substituting the
definition of \(r^w\) and the induction hypothesis into the Bellman recursion
of \(\mathcal M_w\) gives
\[
\begin{aligned}
V_{j,t}^{\pi;\mathcal M_w}(x)
&=\sum_{a,s',o'}\pi^t(a\mid h^t)P(s'\mid s,a)Z(o'\mid s',a)\\
&\quad{}\cdot \sum_{i\in N}w_i
 \left[r_i(s,a,s')+\gamma V_{i,t+1}^{\pi;\mathcal M}(x')\right] =\sum_{i\in N}w_iV_{i,t}^{\pi;\mathcal M}(x).
\end{aligned}
\]
The last equality moves the finite weighted sum outside.  Each remaining term
is the Bellman recursion for agent \(i\) in \(\mathcal M\).  Backward
induction therefore proves that \(V_{j,t}^{\pi;\mathcal M_w}(x)
=\sum_{i\in N}w_iV_{i,t}^{\pi;\mathcal M}(x)\).

For the symmetry inclusion, let \(g\in G_{\mathcal M}^{\mathrm{eval}}\) and
suppose \(w_{\sigma_g(i)}=w_i\) for every \(i\).  Apply the value identity
above to policy \(g\cdot\pi\) and input \(\tau_g^t x\).  Reindexing by
\(k=\sigma_g(i)\) and using weight preservation and policy-evaluation
symmetry in \(\mathcal M\) then gives:
\[
\begin{aligned}
V_{\sigma_g(j),t}^{g\cdot\pi;\mathcal M_w}(\tau_g^t x)
&=\sum_{i\in N}w_{\sigma_g(i)}
 V_{\sigma_g(i),t}^{g\cdot\pi;\mathcal M}(\tau_g^t x)\\
&=\sum_{i\in N}w_iV_{i,t}^{\pi;\mathcal M}(x)
=V_{j,t}^{\pi;\mathcal M_w}(x).
\end{aligned}
\]
Because this holds for every policy, agent, time step, and evaluation input,
we get \(g\in G_{\mathcal M_w}^{\mathrm{eval}}\).

For the componentwise inclusion, let a weight-preserving
\(g\in\AutEnv(\mathcal M)\) be given.  It preserves the transition and
observation distributions copied to \(\mathcal M_w\).  Reindexing the
aggregate reward and using reward preservation gives the equations, as follows:
\[
\begin{aligned}
r^w(\phi_g^{\mathcal S}(s),g\cdot a,\phi_g^{\mathcal S}(s'))
&=\sum_{i\in N}w_{\sigma_g(i)}
r_{\sigma_g(i)}(\phi_g^{\mathcal S}(s),g\cdot a,
                 \phi_g^{\mathcal S}(s'))\\
&=\sum_{i\in N}w_ir_i(s,a,s')
=r^w(s,a,s'),
\end{aligned}
\]
which proves \(g\in\AutEnv(\mathcal M_w)\).
Appendix~\ref{app:cooperative-proofs} gives the complete proof of
Theorem~\ref{thm:weighted-aggregation}.

\paragraph{Equality of weights within structural agent types.}
Theorem~\ref{thm:weighted-aggregation} gives
\(w_{\sigma_g(i)}=w_i\) for every \(i\in N\) as a sufficient condition for a
policy-evaluation symmetry \(g\) of \(\mathcal M\) to be inherited by
\(\mathcal M_w\).  Requiring this condition for every
\(g\in G_{\mathcal M}^{\mathrm{eval}}\) is equivalent to assigning equal
weights to agents in the same structural agent type.  It does not require
equal weights across different types, so all agents must have equal weights only
when \(\mathcal M\) is structurally homogeneous.

If \(w_i = w_j\) for agents \(i\) and \(j\) of the same structural
type in \(\mathcal M\), every policy-evaluation symmetry of
\(\mathcal M\) is inherited by \(\mathcal M_w\), so
\(G_{\mathcal M}^{\mathrm{eval}}
\leq G_{\mathcal M_w}^{\mathrm{eval}}\). Consequently, we get the following partition refinement:
\(
\mathcal P(G_{\mathcal M}^{\mathrm{eval}})
\preceq\mathcal P(G_{\mathcal M_w}^{\mathrm{eval}}).
\)
Because \(\preceq\) means finer than or equal to, every structural agent type
of \(\mathcal M\) is contained in a structural agent type of
\(\mathcal M_w\).  Thus, the aggregate common reward cannot separate agents that share a type
in \(\mathcal M\). Rather, it can only preserve that type or merge it with other
types.  Under this condition, \(\mathcal M_w\) has no more structural agent
types than \(\mathcal M\).  Thus, if \(\mathcal M\) is structurally
homogeneous, then \(\mathcal M_w\) remains structurally homogeneous.

If some agents \(i\) and \(j\) belonging to the same structural agent type in
\(\mathcal M\) receive different weights, Theorem~\ref{thm:weighted-aggregation}
no longer guarantees that a policy-evaluation symmetry of \(\mathcal M\)
mapping \(i\) to \(j\) remains for
\(\mathcal M_w\). If no policy-evaluation symmetry of \(\mathcal M_w\) maps
\(i\) to \(j\), then these agents belong to different structural agent types
in \(\mathcal M_w\). Case~1 in the proof of
Proposition~\ref{prop:supp-aggregation-order} in Appendix~\ref{app:cooperative-proofs} provides a two-agent example
of this splitting.

Conversely, reward aggregation can turn a type-compatible transformation \(g\)
that is not a policy-evaluation symmetry of \(\mathcal M\) into one of
\(\mathcal M_w\). Its agent permutation \(\sigma_g\) may then map
agents belonging to different structural agent types in \(\mathcal M\) into
the same structural agent type in \(\mathcal M_w\). The two-agent examples
in Cases~2 and~3 of the proof of Proposition~\ref{prop:supp-aggregation-order}
demonstrate this merging with equal and unequal weights, respectively.

Therefore, equality of weights within every structural agent type in
\(\mathcal M\) guarantees that
\(
\mathcal P(G_{\mathcal M}^{\mathrm{eval}})
\preceq
\mathcal P(G_{\mathcal M_w}^{\mathrm{eval}}).
\)
Across unrestricted weight assignments, however, there is no universal
refinement relation between these two partitions.

\paragraph{Consequences under the initial distribution.}
Because the two models share the same initial distribution, substituting the
value identity from Theorem~\ref{thm:weighted-aggregation} into the definition of expected return under the initial
distribution in Equation~\eqref{eq:expected-return} gives the following
identity without any symmetry assumption
(see Proposition~\ref{prop:supp-weighted-return}):
\[
J_j^{\mathcal M_w}(\pi)=\sum_{i\in N}w_iJ_i^{\mathcal M}(\pi).
\]
Let \(g\in G_{\mathcal M}^{\mathrm{eval}}\) preserve the weights.
Theorem~\ref{thm:weighted-aggregation} ensures that
\(g\in G_{\mathcal M_w}^{\mathrm{eval}}\). If \(g\) also preserves the
initial distribution, Proposition~\ref{prop:supp-weighted-return} gives
the following expected-return identity:
\[
J_{\sigma_g(j)}^{\mathcal M_w}(g\cdot\pi)
=J_j^{\mathcal M_w}(\pi)
\qquad\text{for every }j\in N.
\]

Figure~\ref{fig:symmetry-hierarchy} summarizes the hierarchy of key concepts and results introduced from Section~\ref{sec:policy-evaluation} to Section~\ref{sec:cooperative}.

\section{Comparisons to Prior Symmetry Frameworks}
\label{sec:comparisons}

We compare two prior frameworks with policy-evaluation symmetry, the details of which are given in Appendix~\ref{app:comparisons}.

\paragraph{Counting and Isomorphic Symmetry.}
In \citet{BraunGL022}'s partitioned Dec-POMDP framework, isomorphic symmetry
uses identical factor templates for agents in each partition part.
Under the grounding assumptions in Section~\ref{app:lifted-isomorphism},
exchanging agents within a part together with their action and observation
assignments preserves the counts in that part and only reorders equal factor
instances. The grounded transition, common-reward, and observation functions
are therefore unchanged, which gives counting symmetry. When these exchanges
are represented by the admissible typed embedding specified there, they are
componentwise environment automorphisms and hence policy-evaluation
symmetries. The lifted partition therefore
refines the structural agent partition. However, the two-agent coordination
example we showed in Appendix~\ref{app:lifted-isomorphism} has counting and policy-evaluation symmetry but lacks the stated repeated additive reward factorization for the same partition with one part.
Policy-evaluation symmetry alone does not guarantee this repeated additive
factorization. Applying \citet{BraunGL022}'s compact-representation and
agent-number tractability results requires their additional factorization
and computational assumptions.

\paragraph{Expected Return Symmetry.}
An expected return symmetry (ERS) preserves the self-play-optimal Boltzmann
policy set at a fixed positive temperature~\citep{muglich2025expected}, whereas
policy-evaluation symmetry preserves complete evaluation for every
decentralized policy. On the shared finite-horizon Dec-POMDP subclass and
common admissible transformations,
Proposition~\ref{prop:ers-noncontainment} establishes that neither ERS nor
initial-distribution-preserving policy-evaluation symmetry implies the other
in general. The two non-implications are established using different models.
For a particular model, the two symmetry sets may coincide or one may contain
the other. Section~\ref{app:expected-return} presents the two counterexamples
as evidence for these non-implications and states the additional local
action-value condition linking the notions. ERS fixes agent indices, so it
does not relate distinct agents
or replace the structural agent types defined by our agent orbits.

\section{Discussions and Framework Boundaries}
\label{sec:discussion}

\paragraph{Why does the initial distribution matter?}
The initial distribution specifies how possible starting states are weighted
when an episode begins.  It is therefore part of standard POSG and Dec-POMDP
formulations~\citep{bernstein2002complexity,hansen2004dynamic}.  A related
separation appears in PlaNet's latent state-space model, which distinguishes
an initial latent state from the transition, observation, and reward models
used to generate a rollout~\citep{hafner2019planet}.  Structural causal models offer a useful
motivation: their structural equations describe transition functions, while a distribution
over exogenous variables of structural equations represent background factors that imply substantive characteristics affecting behavior~\citep{pearl2009causal}. Here, the initial distribution is corresponding to the exogenous distribution.

Our framework therefore separates policy evaluation conditioned on a
state--joint-history input from the initial distribution.  Policy-evaluation symmetry compares values at every
corresponding state--joint-history pair and therefore does not necessarily require
preservation of \(\rho_0\).  For any \(g\in\Geval\), changing variables in
the initial-state sum gives the following relation:
\[
J_{\sigma_g(i)}(g\cdot\pi)
=
\sum_{s\in\mathcal S}
\rho_0\bigl(\phi_g^{\mathcal S}(s)\bigr)
V_{i,0}^{\pi}(s,()^N).
\]
Thus, the return-equivariance identity under the initial distribution holds
for the fixed \(\rho_0\) when \(g\in\Gevali\).

\paragraph{Why go beyond componentwise environment automorphisms?}
Componentwise symmetries have concrete uses.  Graph automorphisms of POMDPs
and POSGs support model reduction~\citep{kang2012symmetries}, Dec-POMDP
equivalence mappings support other-play across conventions~\citep{hu2020otherplay},
and Dec-POMDP automorphisms formalize transformations relevant to zero-shot
coordination~\citep{treutlein2021formalism}.  These approaches preserve the
transition, observation, and reward functions separately.  The resulting
componentwise environment automorphisms provide directly checkable sufficient
conditions for policy-evaluation symmetry, but these conditions need not be
necessary.

Proposition~\ref{prop:strict-witness} gives the concrete separation.  Its
agent transformation changes some entries of the transition distribution, so
it fails the componentwise test.  The changed distributions nevertheless
have the same expected successor-state reward and marginalized
next-observation distribution, which are the quantities used by the Bellman
evaluation in that construction.  The policy-evaluation identity therefore
holds for every decentralized policy.  Two agents share a structural agent type if an admissible transformation
connects them and preserves policy evaluation for every joint policy, even
without preserving each model component. Conversely, componentwise preservation remains an additional, potentially finer diagnostic of where agents differ in the environment representation.

Learning mechanisms concern behavioural policies, so it must be considered when talking about agent heterogeneity, which lets us introduce the policy equivariance. For example, exploration noise whose distribution does not respect the action transformation can violate the equivariance of an already equivariant policy. This will not change agent heterogeneity, because policy-evaluation symmetry requires the evaluation identity to hold for every decentralized policy. This makes the agent heterogeneity decided by the evaluation identity independent of the learning mechanism that is applied to obtain the target policies (that geniunely describes the agent heterogeneity). Thus, our policy-evaluation symmetry can capture intrinsic characteristics of an agent to categorize its agent type.

\paragraph{Why extend the framework to Dec-POMDPs and aggregate common
rewards?}
Dec-POMDPs are a long-standing model of cooperative sequential decision-making
under partial observability, in which decentralized agents optimize a common
reward~\citep{bernstein2002complexity}.  A common reward removes differences
between the agents' reward functions, but it does not guarantee
interchangeable action or observation spaces, or symmetric roles in the
transition and observation distributions. Section~\ref{sec:cooperative} therefore shows that shared objectives need not
imply structural homogeneity. This extension helps connect the objective to
physical and environmental perspectives, as deifned by prior work~\citep{bettini2023heterogeneous}, by assessing their structural
significance through one policy-evaluation symmetry criterion.

Social dilemmas~\citep{dawes1980social} illustrate how individually rational behavior can lead to collectively inferior outcomes, even when cooperation could benefit every player. This motivates mechanisms that support cooperation, with social welfare~\citep{roughgarden2002bad} providing a criterion for evaluating the resulting collective benefits. Social welfare is defined as the sum of individual payoffs, which well matches our definition of the aggregate common reward in Section~\ref{sec:cooperative}. Theorem~\ref{thm:weighted-aggregation} and its examples show that replacing individual rewards by an aggregate common reward can preserve, remove, or create policy-evaluation symmetries. This provides a novel perspective from structural agent types, in contrast to previous work that identifies game structures~\citep{leibo2017multiagent}, to understand the effect of maximizing social welfare to social dilemmas.

In cooperative MARL for solving Dec-POMDPs, the centralised training and decentralised execution (CTDE) paradigm~\citep{oliehoek2008optimal} is proposed to collectively optimize a global value function that approcimates the cumulative common rewards. There are a class of algorithms that proposed to represent the gloabl value function as aggregation of individual values, e.g., VDN uses additive decomposition~\citep{sunehag2018value}; SQDDPG and SHAQ use Shapley-based
value decompositions~\citep{wang2020shapley,wang2022shaq}; and QTRAN learns an additive transformation of the joint action-value function that preserves its optimal actions~\citep{son2019qtran}. As those formulations and the learning paradigm are well aligned with each agent \(j\) optimizing \(V_{j,t}^{\pi;\mathcal M_w}(x)=\sum_{i\in N}w_iV_{i,t}^{\pi;\mathcal M}(x)\), Theorem~\ref{thm:weighted-aggregation} motivates a testable hypothesis: \textit{do these algorithms implicitly learn structural agent types, which could be an inductive bias for improving cooperation?}

\paragraph{Why is one connecting symmetry sufficient?}
We use every element of \(\Geval\) to form orbits: two agents \(i\) and \(j\) belong to the same structural agent type
when some \(g\in\Geval\) satisfies \(\sigma_g(i)=j\). For example, let \(N=\{1,2,3\}\) and consider the induced agent permutations \(\sigma_g\) that fulfil the cyclic permutation
\(1\mapsto2\mapsto3\mapsto1\). Repeated applications of these permutations
can map any agent to any other, so the subgroup it generates acts
transitively on \(N\). This subgroup therefore induces the same
single-part orbit partition \(\{N\}\) as \(\operatorname{Sym}(N)\),
although it contains only three of the six possible agent permutations, as contained in \(\operatorname{Sym}(N)\). In contrast, a
different and much stronger condition is that every agent permutation
\(\xi\in\operatorname{Sym}(N)\) be induced by some \(g\in\Geval\) satisfying
\(\sigma_g=\xi\).  This full-permutation condition is the dynamic analogue of
name-irrelevant symmetry in normal-form games~\citep{cao2018symmetric}.  It is
stronger than structural homogeneity, as we defined in Section~\ref{sec:policy-evaluation}.  As explained after
Definition~\ref{def:structural-type}, structural homogeneity requires only
that, for every pair \(i,j\in N\), some \(g\in\Geval\) satisfies
\(\sigma_g(i)=j\).  The induced agent-permutation group
\(\{\sigma_g:g\in\Geval\}\) may therefore be a proper subgroup of
\(\operatorname{Sym}(N)\). By the definition of stuctural partition, the structural agent types of \(\operatorname{Sym}(N)\) may overclaim the exact structural agent types that can be obtained from \(\Geval\).

The above conclusion is important in applications. Suppose the same stochastic model governs sensor disturbances or physical obstacles for symmetric agents. These factors can be described by symmetries not included in \(\Geval\), and different realized outcomes by those extra symmetries do not by themselves intrinsically change the agent structural types. However, separating stable structures from temporary perturbations, or equivalently, determine a proper subgroup of \(\operatorname{Sym}(N)\), would require deeper investigation.

\paragraph{Framework boundaries.}
The present framework assumes finite state, action, and observation spaces, a
finite horizon, empty initial histories, exact expected-value evaluation, and
the complete domain of decentralized behavioral product policies without
shared randomness.  Its structural types are relative to the specified POSG,
the chosen evaluation semantics, and the declared admissible transformation
group.  Because \(\RAdm\) is declared by the modeller, candidate transformations
must be justified independently of the desired heterogeneity conclusion.
Restricting this group can refine the structural partition, while enlarging it
can coarsen the partition. 

We provide sufficient conditions and finite examples to clarify the policy-evaluation symmetry and the resulting agent heterogeneity, but no general symmetry-discovery or verification algorithm is introduced. Also, exact orbits given by policy-evaluation symmetries do not measure degrees of heterogeneity.

Infinite-horizon models and nonlinear or learned reward aggregation require separate formulations.  The
weighted construction is a forward result for fixed linear weights.  It does
not recover individual rewards from a supplied common reward or cover
nonlinear aggregation.  

Policy-evaluation symmetry compares expected values. It can therefore treat two transformations as symmetric even when they induce different distributions over trajectories or returns, provided those differences do not change the expected values.
Distributional evaluation~\citep{bellemare2017distributional} or
risk-sensitive evaluation~\citep{chow2015risk} may further distinguish such
transformations. Finally, our results do not prove the
existence of an optimal equivariant policy, or establish representation, evaluation and sample-complexity improvements. 

All those above boundaries establishes clear routes for further research to improve our framework. Besides, with proper task formalizations, our framework may also inform research on multi-agent world models~\citep{zhang2025marie} and large language models (LLMs)~\citep{wu2024autogen} by separating structural differences between agents.

\newpage
\section*{Acknowledgments}
Jianhong Wang and Jonathan Lawry are supported by the Engineering and Physical Sciences Research Council [Grant Ref: EP/Y028732/1].

\bibliographystyle{ACM-Reference-Format}
\bibliography{main}

@article{aumann1974subjectivity,
  title={Subjectivity and correlation in randomized strategies},
  author={Aumann, Robert J},
  journal={Journal of mathematical Economics},
  volume={1},
  number={1},
  pages={67--96},
  year={1974},
  publisher={Elsevier}
}

@article{bernstein2009policy,
  author       = {Daniel S. Bernstein and
                  Christopher Amato and
                  Eric A. Hansen and
                  Shlomo Zilberstein},
  title        = {Policy Iteration for Decentralized Control of Markov Decision Processes},
  journal      = {J. Artif. Intell. Res.},
  volume       = {34},
  pages        = {89--132},
  year         = {2009},
}

@article{bernstein2002complexity,
  author       = {Daniel S. Bernstein and
                  Robert Givan and
                  Neil Immerman and
                  Shlomo Zilberstein},
  title        = {The Complexity of Decentralized Control of Markov Decision Processes},
  journal      = {Math. Oper. Res.},
  volume       = {27},
  number       = {4},
  pages        = {819--840},
  year         = {2002},
}

@article{cao2018symmetric,
  author       = {Zhigang Cao and
                  Xiaoguang Yang},
  title        = {Symmetric games revisited},
  journal      = {Math. Soc. Sci.},
  volume       = {95},
  pages        = {9--18},
  year         = {2018},
}

@inproceedings{bettini2023heterogeneous,
  author       = {Matteo Bettini and
                  Ajay Shankar and
                  Amanda Prorok},
  editor       = {Noa Agmon and
                  Bo An and
                  Alessandro Ricci and
                  William Yeoh},
  title        = {Heterogeneous Multi-Robot Reinforcement Learning},
  booktitle    = {Proceedings of the 2023 International Conference on Autonomous Agents
                  and Multiagent Systems, {AAMAS} 2023, London, United Kingdom, 29 May
                  2023 - 2 June 2023},
  pages        = {1485--1494},
  publisher    = {{ACM}},
  year         = {2023},
}

@article{bettini2025system,
  author       = {Matteo Bettini and
                  Ajay Shankar and
                  Amanda Prorok},
  title        = {System Neural Diversity: Measuring Behavioral Heterogeneity in Multi-Agent
                  Learning},
  journal      = {J. Mach. Learn. Res.},
  volume       = {26},
  pages        = {163:1--163:27},
  year         = {2025},
}

@inproceedings{hu2026heterogeneity,
  author       = {Tianyi Hu and
                  Zhiqiang Pu and
                  Yuan Wang and
                  Tenghai Qiu and
                  Min Chen and
                  Xin Yu},
  editor       = {Viviana Mascardi and
                  John Thangarajah and
                  Chris Amato and
                  Louise A. Dennis},
  title        = {Heterogeneity in Multi-Agent Reinforcement Learning},
  booktitle    = {Proceedings of the 25th International Conference on Autonomous Agents
                  and Multiagent Systems, {AAMAS} 2026, Paphos, Cyprus, May 25-29, 2026},
  pages        = {2754--2762},
  publisher    = {{ACM}},
  year         = {2026},
}

@inproceedings{hansen2004dynamic,
  author       = {Eric A. Hansen and
                  Daniel S. Bernstein and
                  Shlomo Zilberstein},
  editor       = {Deborah L. McGuinness and
                  George Ferguson},
  title        = {Dynamic Programming for Partially Observable Stochastic Games},
  booktitle    = {Proceedings of the Nineteenth National Conference on Artificial Intelligence,
                  Sixteenth Conference on Innovative Applications of Artificial Intelligence,
                  July 25-29, 2004, San Jose, California, {USA}},
  pages        = {709--715},
  publisher    = {{AAAI} Press / The {MIT} Press},
  year         = {2004},
}

@inproceedings{hafner2019planet,
  author       = {Danijar Hafner and
                  Timothy P. Lillicrap and
                  Ian Fischer and
                  Ruben Villegas and
                  David Ha and
                  Honglak Lee and
                  James Davidson},
  editor       = {Kamalika Chaudhuri and
                  Ruslan Salakhutdinov},
  title        = {Learning Latent Dynamics for Planning from Pixels},
  booktitle    = {Proceedings of the 36th International Conference on Machine Learning,
                  {ICML} 2019, 9-15 June 2019, Long Beach, California, {USA}},
  series       = {Proceedings of Machine Learning Research},
  volume       = {97},
  pages        = {2555--2565},
  publisher    = {{PMLR}},
  year         = {2019},
}

@inproceedings{hao2023boosting,
  author       = {Jianye Hao and
                  Xiaotian Hao and
                  Hangyu Mao and
                  Weixun Wang and
                  Yaodong Yang and
                  Dong Li and
                  Yan Zheng and
                  Zhen Wang},
  title        = {Boosting Multiagent Reinforcement Learning via Permutation Invariant
                  and Permutation Equivariant Networks},
  booktitle    = {The Eleventh International Conference on Learning Representations,
                  {ICLR} 2023, Kigali, Rwanda, May 1-5, 2023},
  publisher    = {OpenReview.net},
  year         = {2023},
}

@article{kang2012symmetries,
  author       = {Byung Kon Kang and
                  Kee{-}Eung Kim},
  title        = {Exploiting symmetries for single- and multi-agent Partially Observable
                  Stochastic Domains},
  journal      = {Artif. Intell.},
  volume       = {182-183},
  pages        = {32--57},
  year         = {2012},
}

@inproceedings{keriven2019universal,
  author       = {Nicolas Keriven and
                  Gabriel Peyr{\'{e}}},
  editor       = {Hanna M. Wallach and
                  Hugo Larochelle and
                  Alina Beygelzimer and
                  Florence d'Alch{\'{e}}{-}Buc and
                  Emily B. Fox and
                  Roman Garnett},
  title        = {Universal Invariant and Equivariant Graph Neural Networks},
  booktitle    = {Advances in Neural Information Processing Systems 32: Annual Conference
                  on Neural Information Processing Systems 2019, NeurIPS 2019, December
                  8-14, 2019, Vancouver, BC, Canada},
  pages        = {7090--7099},
  year         = {2019},
}

@inproceedings{koriche2017constraint,
  author       = {Fr{\'{e}}d{\'{e}}ric Koriche and
                  Sylvain Lagrue and
                  {\'{E}}ric Piette and
                  S{\'{e}}bastien Tabary},
  editor       = {Carles Sierra},
  title        = {Constraint-Based Symmetry Detection in General Game Playing},
  booktitle    = {Proceedings of the Twenty-Sixth International Joint Conference on
                  Artificial Intelligence, {IJCAI} 2017, Melbourne, Australia, August
                  19-25, 2017},
  pages        = {280--287},
  publisher    = {ijcai.org},
  year         = {2017},
}

@inproceedings{hu2020otherplay,
  author       = {Hengyuan Hu and
                  Adam Lerer and
                  Alex Peysakhovich and
                  Jakob N. Foerster},
  title        = {"Other-Play" for Zero-Shot Coordination},
  booktitle    = {Proceedings of the 37th International Conference on Machine Learning,
                  {ICML} 2020, 13-18 July 2020, Virtual Event},
  series       = {Proceedings of Machine Learning Research},
  volume       = {119},
  pages        = {4399--4410},
  publisher    = {{PMLR}},
  year         = {2020},
}

@inproceedings{treutlein2021formalism,
  author       = {Johannes Treutlein and
                  Michael Dennis and
                  Caspar Oesterheld and
                  Jakob N. Foerster},
  editor       = {Marina Meila and
                  Tong Zhang},
  title        = {A New Formalism, Method and Open Issues for Zero-Shot Coordination},
  booktitle    = {Proceedings of the 38th International Conference on Machine Learning,
                  {ICML} 2021, 18-24 July 2021, Virtual Event},
  series       = {Proceedings of Machine Learning Research},
  volume       = {139},
  pages        = {10413--10423},
  publisher    = {{PMLR}},
  year         = {2021},
}

@inproceedings{tewolde2025computing,
  author       = {Emanuel Tewolde and
                  Brian Hu Zhang and
                  Caspar Oesterheld and
                  Tuomas Sandholm and
                  Vincent Conitzer},
  editor       = {Toby Walsh and
                  Julie Shah and
                  Zico Kolter},
  title        = {Computing Game Symmetries and Equilibria That Respect Them},
  booktitle    = {Thirty-Ninth {AAAI} Conference on Artificial Intelligence, Thirty-Seventh
                  Conference on Innovative Applications of Artificial Intelligence,
                  Fifteenth Symposium on Educational Advances in Artificial Intelligence,
                  {AAAI} 2025, Philadelphia, PA, USA, February 25 - March 4, 2025},
  pages        = {14148--14157},
  publisher    = {{AAAI} Press},
  year         = {2025},
}

@book{judson1994abstract,
  title={Abstract algebra: theory and applications},
  author={Judson, Thomas W and Beezer, Robert Arnold},
  volume={20},
  year={1994},
  publisher={PWS Publishing Company Boston}
}

@inproceedings{vanderpol2022multiagent,
  author       = {Elise van der Pol and
                  Herke van Hoof and
                  Frans A. Oliehoek and
                  Max Welling},
  title        = {Multi-Agent {MDP} Homomorphic Networks},
  booktitle    = {The Tenth International Conference on Learning Representations, {ICLR}
                  2022, Virtual Event, April 25-29, 2022},
  publisher    = {OpenReview.net},
  year         = {2022},
}

@inproceedings{vanderpol2020mdp,
  author       = {Elise van der Pol and
                  Daniel E. Worrall and
                  Herke van Hoof and
                  Frans A. Oliehoek and
                  Max Welling},
  editor       = {Hugo Larochelle and
                  Marc'Aurelio Ranzato and
                  Raia Hadsell and
                  Maria{-}Florina Balcan and
                  Hsuan{-}Tien Lin},
  title        = {{MDP} Homomorphic Networks: Group Symmetries in Reinforcement Learning},
  booktitle    = {Advances in Neural Information Processing Systems 33: Annual Conference
                  on Neural Information Processing Systems 2020, NeurIPS 2020, December
                  6-12, 2020, virtual},
  year         = {2020},
}

@inproceedings{BraunGL022,
  author       = {Tanya Braun and
                  Marcel Gehrke and
                  Florian Lau and
                  Ralf M{\"{o}}ller},
  editor       = {James Cussens and
                  Kun Zhang},
  title        = {Lifting in multi-agent systems under uncertainty},
  booktitle    = {Uncertainty in Artificial Intelligence, Proceedings of the Thirty-Eighth
                  Conference on Uncertainty in Artificial Intelligence, {UAI} 2022,
                  1-5 August 2022, Eindhoven, The Netherlands},
  series       = {Proceedings of Machine Learning Research},
  volume       = {180},
  pages        = {233--243},
  publisher    = {{PMLR}},
  year         = {2022},
}

@inproceedings{muglich2025expected,
  author       = {Darius Muglich and
                  Johannes Forkel and
                  Elise van der Pol and
                  Jakob Nicolaus Foerster},
  title        = {Expected Return Symmetries},
  booktitle    = {The Thirteenth International Conference on Learning Representations,
                  {ICLR} 2025, Singapore, April 24-28, 2025},
  publisher    = {OpenReview.net},
  year         = {2025},
}

@article{pearl2009causal,
author = {Judea Pearl},
title = {{Causal inference in statistics: An overview}},
volume = {3},
journal = {Statistics Surveys},
number = {none},
publisher = {Amer. Statist. Assoc., the Bernoulli Soc., the Inst. Math. Statist., and the Statist. Soc. Canada},
pages = {96 -- 146},
}

@inproceedings{sunehag2018value,
  author       = {Peter Sunehag and
                  Guy Lever and
                  Audrunas Gruslys and
                  Wojciech Marian Czarnecki and
                  Vin{\'{\i}}cius Flores Zambaldi and
                  Max Jaderberg and
                  Marc Lanctot and
                  Nicolas Sonnerat and
                  Joel Z. Leibo and
                  Karl Tuyls and
                  Thore Graepel},
  editor       = {Elisabeth Andr{\'{e}} and
                  Sven Koenig and
                  Mehdi Dastani and
                  Gita Sukthankar},
  title        = {Value-Decomposition Networks For Cooperative Multi-Agent Learning
                  Based On Team Reward},
  booktitle    = {Proceedings of the 17th International Conference on Autonomous Agents
                  and MultiAgent Systems, {AAMAS} 2018, Stockholm, Sweden, July 10-15,
                  2018},
  pages        = {2085--2087},
  publisher    = {International Foundation for Autonomous Agents and Multiagent Systems
                  Richland, SC, {USA} / {ACM}},
  year         = {2018},
}

@article{molnar2021asymmetry,
  title={Asymmetry underlies stability in power grids},
  author={Molnar, Ferenc and Nishikawa, Takashi and Motter, Adilson E},
  journal={Nature communications},
  volume={12},
  number={1},
  pages={1457},
  year={2021},
  publisher={Nature Publishing Group UK London}
}

@inproceedings{schutt2017schnet,
  author       = {Kristof Sch{\"{u}}tt and
                  Pieter{-}Jan Kindermans and
                  Huziel Enoc Sauceda Felix and
                  Stefan Chmiela and
                  Alexandre Tkatchenko and
                  Klaus{-}Robert M{\"{u}}ller},
  editor       = {Isabelle Guyon and
                  Ulrike von Luxburg and
                  Samy Bengio and
                  Hanna M. Wallach and
                  Rob Fergus and
                  S. V. N. Vishwanathan and
                  Roman Garnett},
  title        = {SchNet: {A} continuous-filter convolutional neural network for modeling
                  quantum interactions},
  booktitle    = {Advances in Neural Information Processing Systems 30: Annual Conference
                  on Neural Information Processing Systems 2017, December 4-9, 2017,
                  Long Beach, CA, {USA}},
  pages        = {991--1001},
  year         = {2017},
}

@inproceedings{wang2021voltage,
  author       = {Jianhong Wang and
                  Wangkun Xu and
                  Yunjie Gu and
                  Wenbin Song and
                  Tim C. Green},
  editor       = {Marc'Aurelio Ranzato and
                  Alina Beygelzimer and
                  Yann N. Dauphin and
                  Percy Liang and
                  Jennifer Wortman Vaughan},
  title        = {Multi-Agent Reinforcement Learning for Active Voltage Control on Power
                  Distribution Networks},
  booktitle    = {Advances in Neural Information Processing Systems 34: Annual Conference
                  on Neural Information Processing Systems 2021, NeurIPS 2021, December
                  6-14, 2021, virtual},
  pages        = {3271--3284},
  year         = {2021},
}

@inproceedings{wang2022shaq,
  author       = {Jianhong Wang and
                  Yuan Zhang and
                  Yunjie Gu and
                  Tae{-}Kyun Kim},
  editor       = {Sanmi Koyejo and
                  S. Mohamed and
                  A. Agarwal and
                  Danielle Belgrave and
                  K. Cho and
                  A. Oh},
  title        = {{SHAQ:} Incorporating Shapley Value Theory into Multi-Agent Q-Learning},
  booktitle    = {Advances in Neural Information Processing Systems 35: Annual Conference
                  on Neural Information Processing Systems 2022, NeurIPS 2022, New Orleans,
                  LA, USA, November 28 - December 9, 2022},
  year         = {2022},
}

@inproceedings{wang2020shapley,
  author       = {Jianhong Wang and
                  Yuan Zhang and
                  Tae{-}Kyun Kim and
                  Yunjie Gu},
  title        = {Shapley Q-Value: {A} Local Reward Approach to Solve Global Reward
                  Games},
  booktitle    = {The Thirty-Fourth {AAAI} Conference on Artificial Intelligence, {AAAI}
                  2020, The Thirty-Second Innovative Applications of Artificial Intelligence
                  Conference, {IAAI} 2020, The Tenth {AAAI} Symposium on Educational
                  Advances in Artificial Intelligence, {EAAI} 2020, New York, NY, USA,
                  February 7-12, 2020},
  pages        = {7285--7292},
  publisher    = {{AAAI} Press},
  year         = {2020},
}

@inproceedings{son2019qtran,
  author       = {Kyunghwan Son and
                  Daewoo Kim and
                  Wan Ju Kang and
                  David Hostallero and
                  Yung Yi},
  editor       = {Kamalika Chaudhuri and
                  Ruslan Salakhutdinov},
  title        = {{QTRAN:} Learning to Factorize with Transformation for Cooperative
                  Multi-Agent Reinforcement Learning},
  booktitle    = {Proceedings of the 36th International Conference on Machine Learning,
                  {ICML} 2019, 9-15 June 2019, Long Beach, California, {USA}},
  series       = {Proceedings of Machine Learning Research},
  volume       = {97},
  pages        = {5887--5896},
  publisher    = {{PMLR}},
  year         = {2019},
}

@inproceedings{bellemare2017distributional,
  author       = {Marc G. Bellemare and
                  Will Dabney and
                  R{\'{e}}mi Munos},
  editor       = {Doina Precup and
                  Yee Whye Teh},
  title        = {A Distributional Perspective on Reinforcement Learning},
  booktitle    = {Proceedings of the 34th International Conference on Machine Learning,
                  {ICML} 2017, Sydney, NSW, Australia, 6-11 August 2017},
  series       = {Proceedings of Machine Learning Research},
  volume       = {70},
  pages        = {449--458},
  publisher    = {{PMLR}},
  year         = {2017},
}

@inproceedings{chow2015risk,
  author       = {Yinlam Chow and
                  Aviv Tamar and
                  Shie Mannor and
                  Marco Pavone},
  editor       = {Corinna Cortes and
                  Neil D. Lawrence and
                  Daniel D. Lee and
                  Masashi Sugiyama and
                  Roman Garnett},
  title        = {Risk-Sensitive and Robust Decision-Making: a CVaR Optimization Approach},
  booktitle    = {Advances in Neural Information Processing Systems 28: Annual Conference
                  on Neural Information Processing Systems 2015, December 7-12, 2015,
                  Montreal, Quebec, Canada},
  pages        = {1522--1530},
  year         = {2015},
}

@article{wu2024autogen,
  title={Autogen: Enabling next-gen llm applications via multi-agent conversation},
  author={Wu, Qingyun and Bansal, Gagan and Zhang, Jieyu and Wu, Yiran and Li, Beibin and Zhu, Erkang and Jiang, Li and Zhang, Xiaoyun and Zhang, Shaokun and Liu, Jiale and others},
  journal={arXiv preprint arXiv:2308.08155},
  year={2023}
}

@article{zhang2025marie,
  author       = {Yang Zhang and
                  Chenjia Bai and
                  Bin Zhao and
                  Junchi Yan and
                  Xiu Li and
                  Xuelong Li},
  title        = {Decentralized Transformers with Centralized Aggregation are Sample-Efficient
                  Multi-Agent World Models},
  journal      = {Trans. Mach. Learn. Res.},
  volume       = {2025},
  year         = {2025},
}

@inproceedings{leibo2017multiagent,
  author       = {Joel Z. Leibo and
                  Vin{\'{\i}}cius Flores Zambaldi and
                  Marc Lanctot and
                  Janusz Marecki and
                  Thore Graepel},
  editor       = {Kate Larson and
                  Michael Winikoff and
                  Sanmay Das and
                  Edmund H. Durfee},
  title        = {Multi-agent Reinforcement Learning in Sequential Social Dilemmas},
  booktitle    = {Proceedings of the 16th Conference on Autonomous Agents and MultiAgent
                  Systems, {AAMAS} 2017, S{\~{a}}o Paulo, Brazil, May 8-12, 2017},
  pages        = {464--473},
  publisher    = {{ACM}},
  year         = {2017},
}

@article{dawes1980social,
  title={Social dilemmas.},
  author={Dawes, Robyn M},
  journal={Annual review of psychology},
  year={1980},
  publisher={Annual Reviews}
}

@article{roughgarden2002bad,
  author       = {Tim Roughgarden and
                  {\'{E}}va Tardos},
  title        = {How bad is selfish routing?},
  journal      = {J. {ACM}},
  volume       = {49},
  number       = {2},
  pages        = {236--259},
  year         = {2002},
}

@article{oliehoek2008optimal,
  author       = {Frans A. Oliehoek and
                  Matthijs T. J. Spaan and
                  Nikos Vlassis},
  title        = {Optimal and Approximate Q-value Functions for Decentralized POMDPs},
  journal      = {CoRR},
  volume       = {abs/1111.0062},
  year         = {2011},
}

\appendix

\clearpage
\newpage
\section{Detailed Comparison of Heterogeneity Categories}
\label{app:heterogeneity-comparison}

Table~\ref{tab:heterogeneity-comparison} compares prior categories and measures
of agent heterogeneity to different levels of our framework.  The two approaches ask
different questions.  Existing taxonomies primarily identify where agents
differ, such as in their physical capabilities, observations, dynamics,
objectives, or selected policies.  Our framework instead asks whether an
admissible transformation can exchange the agents while preserving complete
policy evaluation.  The correspondences are therefore conditional
interpretations rather than equivalences between definitions.

The rows concerning physical properties, observations, transitions, and
objectives share one logical direction.  For agents \(i,j\in N\),
Theorem~\ref{thm:componentwise-certificate} gives
\[
\bigl[\text{some }g\in\AutEnv(\mathcal M)
\text{ satisfies }\sigma_g(i)=j\bigr]
\quad\Longrightarrow\quad
i\sim_{\mathrm{eval}}j.
\]
The converse is false in general by
Proposition~\ref{prop:strict-witness}.  Consequently, componentwise
preservation supplies a sufficient test for agents to share a structural
agent type, but failure of a componentwise preservation equation for one
candidate transformation does not establish that their structural types differ.
Another admissible transformation may exchange the same agents.  By contrast, if
no admissible transformation maps \(i\) to \(j\), then no policy-evaluation
symmetry can map them, and they necessarily have different structural agent
types.

\subsection{Physical Heterogeneity}
Bettini et al.~\citep{bettini2023heterogeneous} describe physical
heterogeneity through differences in hardware or physical constraints, such
as sensors, actuators, and motion constraints.  Our framework can classify
such differences only through their representation in the POSG.  Different
sensors may appear in the observation spaces or observation distribution,
different actuators may appear in the action spaces or transition
distribution, and motion constraints may appear in the state and transition
distribution.  If action or observation spaces have different cardinalities,
the required typed bijection does not exist.  Equal cardinality permits a
bijection between differently named spaces, but it does not by itself place
the resulting complete transformation in the application-specified admissible
group.  Thus, the formal comparison concerns the modeled consequences of a
physical difference rather than hardware identity itself.  A physical
property that is not represented in the POSG lies outside the structural
classification.

\subsection{Observation Heterogeneity}
Hu et al.~\citep{hu2026heterogeneity} define observation heterogeneity through
differences in agents' observation spaces or local observation functions.
Our closest test combines the typed observation maps
\(\phi_{g,i}^{\mathcal O}\) with the joint observation-preservation condition
in Equation~\eqref{eq:component-Z}.  The comparison is made after the state,
action, and observation labels have been transformed.  Hence differently
named observations need not indicate different structural agent types.  To
compare directly with Hu et al.'s agent-local formulation, an application
must also specify compatible timing and conditioning conventions and explain
how its local observation functions induce the joint observation distribution
\(Z\).  If Equation~\eqref{eq:component-Z} fails, that particular
componentwise environment automorphism is excluded, but a policy-evaluation
symmetry may still relate the agents.

\subsection{Response- and Effect-Transition Heterogeneity}
Hu et al.~\citep{hu2026heterogeneity} distinguish two locations at which
agents may differ in the transition dynamics.  Their agent-level model
decomposes the global state as
\[
\widehat s=(s_1,\ldots,s_{|N|},s_E)
\in
\left(\prod_{k\in N}\mathcal S_k\right)\times\mathcal S_E,
\]
where \(s_k\) is agent \(k\)'s local state and \(s_E\) contains the remaining
environmental state.  Write
\(\widehat a=(a_1,\ldots,a_{|N|})\) for the joint action.
Response-transition heterogeneity concerns the local
transition law \(T^i(\cdot\mid\widehat s,\widehat a)\): under corresponding
global states and joint actions, it asks whether agents have different
distributions over their own next local states.  Effect-transition
heterogeneity instead concerns \(T^{-i}\), the collection of transition laws
for the state components other than agent \(i\), viewed as functions of
\((s_i,a_i)\) as well as the remaining state and action components.  It asks
whether different agents' states or actions influence the rest of the system
in different ways.

The core POSG in Section~\ref{sec:setting} does not assume this agent-level
decomposition.  It specifies an undivided state space \(\mathcal S\) and one
global transition distribution \(P(s'\mid s,a),\) without canonically assigning coordinates of \(s\) to individual agents or attributing parts of the transition law to particular update mechanisms.
Consequently, a difference in an agent's own next-state distribution and a
difference in its influence on other next-state components both appear as
differences in the same global distribution.  Equation~\eqref{eq:component-P}
can test whether that entire distribution is preserved by a candidate type-compatible
transformation.  If the equation fails, however, the undivided representation
does not reveal whether the failure arose from response dynamics, effect
dynamics, or both.  Thus, the two categories have the same closest
counterpart in our core model only in this representational sense.  They are
not treated as equivalent concepts.

To distinguish them, an application may additionally declare a decomposition
\(\mathcal S=(\prod_{k\in N}\mathcal S_k)\times\mathcal S_E\), together with a
complete collection of local transition factors and their dependency scopes.
A candidate transformation must then map each local state component and transition
factor to the corresponding component and factor for the target agent.  A
response-preservation test compares the factors that generate \(s_i'\) and
\(s_{\sigma_g(i)}'\).  An effect-preservation test instead compares the
factors for the remaining agents and environmental state insofar as those
factors depend on \((s_i,a_i)\) and
\((s_{\sigma_g(i)},a_{\sigma_g(i)})\), respectively.  This additional
structure is application-specific and, in general, cannot be recovered
uniquely from \(P\) alone.  For example, two robots exhibit a response
difference if the same drive command under corresponding conditions produces
different distributions over their own next velocities.  They exhibit an
effect difference if corresponding pushing actions produce different
distributions over the next position of a shared object.  Without declared
velocity and object-state components, both differences appear only as changes
in \(P(s'\mid s,a)\).

These factor-level tests identify where a difference in the dynamics occurs.
They do not provide a second definition of structural agent type.  Structural
agent types remain the agent orbits of \(G_{\mathcal M}^{\mathrm{eval}}\)
relative to the resulting admissible transformation group.  Restricting that group
to transformations that respect the declared factorization can refine the
resulting orbit partition.  Moreover, failure of a response- or
effect-preservation test for one candidate transformation excludes that candidate
componentwise symmetry, but it does not by itself prove that two agents have
different structural agent types: another admissible policy-evaluation
symmetry may still exchange them.

\subsection{Objective Heterogeneity}
Hu et al.~\citep{hu2026heterogeneity} compare agents' conditional reward
distributions, while
Bettini et al.~\citep{bettini2023heterogeneous} distinguish behavior under a
shared objective from behavior under different objectives.  The closest
componentwise test under our deterministic transition-reward convention is
the exact reward-preservation condition in
Equation~\eqref{eq:component-r}.  It compares the reward functions after the
agent, state, and action labels have been transformed, rather than requiring
the functions to be identical under their source labels.  Failure of this
condition excludes that particular componentwise environment automorphism but
does not, by itself, establish different structural agent types.

Section~\ref{sec:cooperative} also shows that a common reward function does
not imply structural homogeneity.  In a Dec-POMDP, all agents
have equal value coordinates under the same policy and state--joint-history
input, but they need not be exchangeable in their action spaces, observation
spaces, or roles in the transition and observation distributions.  Conversely,
forming an aggregate common reward can preserve, remove, or create
policy-evaluation symmetries, depending on the individual rewards and
aggregation weights.  Objective heterogeneity and structural heterogeneity
are therefore related but distinct properties.

\subsection{Behavioral and Policy Heterogeneity}
The policy-output aspect of behavioral and policy heterogeneity concerns a
selected policy rather than only the underlying game.  Suppose
\(H\leq G_{\mathcal M}^{\mathrm{eval}}\).  A selected policy \(\pi\) is
\(H\)-equivariant exactly when, for every \(g\in H\), agent \(i\), decision
time step \(t<T\), local history \(h_i^t\), and action \(a_i\),
\[
\pi_{\sigma_g(i)}^t
\bigl(\phi_{g,i}^{\mathcal A}(a_i)\mid
\Phi_{g,i}^{\mathcal H,t}(h_i^t)\bigr)
=
\pi_i^t(a_i\mid h_i^t).
\]
After the prescribed history and action transformations, the two local rules
therefore assign the same probabilities to corresponding actions.  This is
the symmetry-aware condition for behavioral homogeneity within each agent
orbit of \(H\).  The policy tables may look different under their source
labels and still satisfy it.

For example, consider two agents in mirrored roles.  One agent may turn right
after observing an obstacle on its left, while the corresponding rule for the
other agent is to turn left after observing an obstacle on its right.  The
rules are equivariant if the game symmetry exchanges the agents and the
left--right labels.  If a selected policy fails the displayed equality for
this transformation, it breaks that particular game symmetry and expresses a
policy-relative behavioral difference.  This does not change the structural
agent types, which are defined from complete policy evaluation for every
policy.  Nor does failure for one transformation prove that the agents are
unrelated at the policy level, because another policy-evaluation symmetry may
both preserve the selected policy and exchange them.

The policy-heterogeneity definition of Hu et
al.~\citep{hu2026heterogeneity} covers more than differences between policy
outputs: it also includes differences in action or observation spaces.  In
our framework, the existence of the
necessary bijections is settled first by the type-compatibility and bijectivity requirements.
Policy equivariance then compares the decision rules after those bijections
have been applied.  Bettini et al.~\citep{bettini2023heterogeneous} use the
possibility of distinct policy outputs on the same input to characterize
agents with different behavioral models.  They do not, however, restrict
their analysis to identical inputs.  Their behavioral-typing analysis also
studies how one shared policy can produce apparently heterogeneous behavior
when agent identifiers or other differences in observations provide different
inputs.  Our equivariance condition addresses another question: whether the
selected local policies agree at histories and actions related by a
policy-evaluation symmetry.  Specifically, it compares the probability
assigned to \(a_i\) at \(h_i^t\) with the probability assigned to
\(\phi_{g,i}^{\mathcal A}(a_i)\) at
\(\Phi_{g,i}^{\mathcal H,t}(h_i^t)\) by agent \(\sigma_g(i)\).  When the
history and action transformations are identities, this reduces to the same-input
comparison.

\subsection{Exact Classes and Graded Differences}
The structural partition \(\Peval(\mathcal M)\) is an exact classification:
two agents share a structural agent type when some policy-evaluation symmetry
exchanges them.  The system neural diversity measure~\citep{bettini2025system} and the
heterogeneity distances of Hu et al.~\citep{hu2026heterogeneity} instead
measure degrees of difference between behavioral or model components.  Such
measures can compare agents within one structural agent type or in different
types when their required domains are comparable, but they do not replace the
orbit partition.  Conversely, orbit membership does not assign a numerical
degree of similarity.

The final row of Table~\ref{tab:heterogeneity-comparison} locates system
neural diversity at the selected-policy level.  Its pairwise distances
compare agents' action distributions on common observation inputs, assuming
shared observation and action spaces~\citep{bettini2025system}.
This comparison does not test preservation of complete policy evaluation
under agent permutations.

\subsection{Summary}
In summary, the transition, observation, and reward preservation conditions
diagnose separately represented parts of the model. Policy-evaluation symmetries define structural agent types for the specified POSG and admissible transformation group. Given these symmetries, policy equivariance determines whether a selected policy is invariant under them. Table~\ref{tab:heterogeneity-comparison} classifies the prior categories according to the level of our framework that they describe. Structural agent types are instead defined by the agent orbits of the policy-evaluation symmetry group.

\section{Proofs for the Core Framework}
\label{app:foundations}

In the following proofs, we define the product spaces for the joint action, joint observation, and joint history at time step \(t\), as follows:
\[
\mathcal A=\prod_{i\in N}\mathcal A_i,
\qquad
\mathcal O=\prod_{i\in N}\mathcal O_i,
\qquad
\mathcal H^t=\prod_{i\in N}\mathcal H_i^t.
\]
For transformations \(g,k\), the product \(gk\) means first applying \(k\) and
then \(g\).  Thus, we have the following composition laws for the agent permutation and the state, action, and observation maps:
\[
\sigma_{gk}=\sigma_g\circ\sigma_k,
\qquad
\phi_{gk}^{\mathcal S}
=\phi_g^{\mathcal S}\circ\phi_k^{\mathcal S},
\]
\[
\phi_{gk,i}^{\mathcal A}
=\phi_{g,\sigma_k(i)}^{\mathcal A}\circ\phi_{k,i}^{\mathcal A},
\qquad
\phi_{gk,i}^{\mathcal O}
=\phi_{g,\sigma_k(i)}^{\mathcal O}\circ\phi_{k,i}^{\mathcal O}.
\]
We write \((h^t,a,o')\) for the joint history obtained by appending
\((a_i,o_i')\) to each local history \(h_i^t\).

\subsection{Induced actions}

\begin{proposition}[Induced group actions]
\label{prop:supp-induced-actions}
The transformations in Equations~\eqref{eq:policy-action}
and~\eqref{eq:value-action} define left group actions of \(\RAdm\) on
\(\Pi_{\mathrm{dec}}\) and \(\mathcal V\), respectively.  Specifically:

\emph{(i) Policies.} For every \(g,k\in\RAdm\) and
\(\pi\in\Pi_{\mathrm{dec}}\), the transformed policy belongs to
\(\Pi_{\mathrm{dec}}\), and
\[
e\cdot\pi=\pi,
\qquad
g\cdot(k\cdot\pi)=(gk)\cdot\pi.
\]
For every \(t<T\), joint history \(h^t\), and joint action \(a\),
\[
(g\cdot\pi)^t(g\cdot a\mid g\cdot h^t)
=\pi^t(a\mid h^t)
=\prod_{i\in N}\pi_i^t(a_i\mid h_i^t).
\]
Consequently, at every fixed joint history the transformed joint action
distribution is still the product of one local distribution per agent.  The
deterministic transformation introduces no shared policy-randomization variable
and no additional conditional correlation between agents' actions.

\emph{(ii) Value profiles.} For every \(g,k\in\RAdm\) and
\(V\in\mathcal V\), the transformed profile belongs to \(\mathcal V\), and
\[
e\cdot V=V,
\qquad
g\cdot(k\cdot V)=(gk)\cdot V.
\]
\end{proposition}

The value-profile action laws ensure that successive transformations compose
consistently.  They are used in
Proposition~\ref{prop:supp-policy-evaluation-subgroup} to establish that
policy-evaluation symmetries form a subgroup.

\begin{proof}
By the coordinatewise definition of the local-history maps and the
composition laws for the action and observation maps, we obtain the following equation:
\[
\Phi_{gk,i}^{\mathcal H,t}
=
\Phi_{g,\sigma_k(i)}^{\mathcal H,t}
\circ\Phi_{k,i}^{\mathcal H,t},
\]
where \(\sigma_k(i)\) identifies the agent whose coordinate contains the
local history after applying \(k\).  At the initial time step \(t=0\), both
sides fix the empty history.

The joint-history transformation collects these local-history maps in the
permuted agent coordinates.  Using
\(\sigma_{gk}=\sigma_g\circ\sigma_k\), for every \(i\in N\) we have that:
\[
\begin{aligned}
&[g\cdot(k\cdot h^t)]_{\sigma_{gk}(i)}\\
&\quad=\Phi_{g,\sigma_k(i)}^{\mathcal H,t}
  \bigl(\Phi_{k,i}^{\mathcal H,t}(h_i^t)\bigr)\\
&\quad=\Phi_{gk,i}^{\mathcal H,t}(h_i^t)
=[(gk)\cdot h^t]_{\sigma_{gk}(i)}.
\end{aligned}
\]
Since \(\sigma_{gk}\) is a permutation, this proves that:
\[
(gk)\cdot h^t=g\cdot(k\cdot h^t).
\]
Finally, combining this identity with
\(\phi_{gk}^{\mathcal S}=\phi_g^{\mathcal S}\circ\phi_k^{\mathcal S}\)
and the definition
\(\tau_g^t(s,h^t)=(\phi_g^{\mathcal S}(s),g\cdot h^t)\) gives that:
\[
\tau_{gk}^t=\tau_g^t\circ\tau_k^t.
\]
The identity acts trivially, and the inverse coordinate maps undo each
transformation.

For policies, put \(a'=g\cdot a\) and \(h'=g\cdot h^t\). Then, we have that:
\[
\begin{aligned}
(g\cdot\pi)^t(a'\mid h')
&=\prod_{j\in N}(g\cdot\pi)_j^t(a'_j\mid h'_j)\\
&=\prod_{i\in N}(g\cdot\pi)_{\sigma_g(i)}^t
  (a'_{\sigma_g(i)}\mid h'_{\sigma_g(i)})\\
&=\prod_{i\in N}\pi_i^t(a_i\mid h_i^t)
=\pi^t(a\mid h^t).
\end{aligned}
\]
Each transformed local rule depends only on the corresponding local history.
It is normalized because its probabilities are obtained by bijectively
reindexing the source agent's actions.  The pointwise policy definition and the
coordinate composition laws give
\(g\cdot(k\cdot\pi)=(gk)\cdot\pi\) and \(e\cdot\pi=\pi\).

For the action on \(\mathcal V\), recall that this is the space of real-valued
function profiles with zero terminal entries.  Policy evaluation produces a
particular profile \(V^\pi\) in this space.  Fix an arbitrary
\(V\in\mathcal V\).  Since \(\sigma_g\) and \(\tau_g^t\)
are bijective, Equation~\eqref{eq:value-action} assigns a unique real value
to every transformed coordinate.  The time index is unchanged, so terminal
entries remain zero and \(g\cdot V\in\mathcal V\).

For any \(g,k\in\RAdm\), \(i\in N\), \(0\leq t\leq T\), and
\(x_t\in\mathcal X_t\), applying Equation~\eqref{eq:value-action} twice gives that:
\[
\begin{aligned}
&[g\cdot(k\cdot V)]_{\sigma_g(\sigma_k(i)),t}
  \bigl(\tau_g^t(\tau_k^t x_t)\bigr)\\
&\quad=(k\cdot V)_{\sigma_k(i),t}(\tau_k^t x_t)\\
&\quad=V_{i,t}(x_t).
\end{aligned}
\]
The first equality applies the definition to the profile \(k\cdot V\)
under \(g\), at the coordinate \((\sigma_k(i),t,\tau_k^t x_t)\).
The second applies it to \(V\) under \(k\).
Applying the combined transformation \(gk\) once gives that:
\[
[(gk)\cdot V]_{\sigma_{gk}(i),t}(\tau_{gk}^t x_t)
=V_{i,t}(x_t).
\]
Because \(\sigma_{gk}=\sigma_g\circ\sigma_k\) and
\(\tau_{gk}^t=\tau_g^t\circ\tau_k^t\), the two calculations evaluate
\(g\cdot(k\cdot V)\) and \((gk)\cdot V\) at the same coordinate.
Bijectivity ensures that these coordinates cover every agent and input
at each time step, proving \(g\cdot(k\cdot V)=(gk)\cdot V\).
For the identity transformation, \(\sigma_e\) and \(\tau_e^t\) are identity
maps, so Equation~\eqref{eq:value-action} gives \(e\cdot V=V\).
This proves the value-profile action laws and completes the proof.
\end{proof}

\subsection{Complete proof for the strictness example}
\label{app:strict-witness}

\begin{proof}[Proof of Proposition~\ref{prop:strict-witness}]
Let \(N=\{1,2\}\), set \(T=2\), choose any \(\gamma\in(0,1]\), and let
\(\mathcal S=\{L,M,H\}\).  Both agents have action and observation sets
\(\{0,1\}\), and the initial state distribution is uniform.  Take
\(\mathcal R_{\mathrm{adm}}(\mathcal M)\) to be the full group of complete
type-compatible transformations of these finite sets.  Let \(g\) exchange the two
agents while fixing the state, action, and observation labels.  Thus,
\[
g\cdot(a_1,a_2)=(a_2,a_1),
\qquad
g\cdot(o_1,o_2)=(o_2,o_1).
\]

For a joint action \(a=(a_1,a_2)\), abbreviated as \(a_1a_2\), make the
transition distribution independent of the current state by setting
\[
P(s'\mid s,a)=p_a(s')
\qquad\text{for every }s\in\mathcal S.
\]
In next-state order \((L,M,H)\), define
\[
\begin{array}{c@{\quad}c@{\qquad}c@{\quad}c}
a&p_a&a&p_a\\
00&(1/2,1/3,1/6)&01&(1/3,1/3,1/3)\\
10&(5/12,1/6,5/12)&11&(1/6,1/3,1/2).
\end{array}
\]
Make the observation distribution depend on the successor state but not on
the joint action:
\[
Z(o'\mid s',a)=q_{s'}(o').
\]
Abbreviating a joint observation \((o_1',o_2')\) as \(o_1'o_2'\), use the
joint-observation order \((00,01,10,11)\) and set
\[
q_L=(3/8,1/4,1/4,1/8),\quad
q_M=(1/4,1/4,1/4,1/4),
\]
\[
q_H=(1/8,1/4,1/4,3/8).
\]
Finally, define
\[
c(L)=0,\qquad c(M)=1,\qquad c(H)=2,
\]
and
\[
r_i(s,a,s')=c(s)+c(s')+2a_i+\mathbf 1\{a_1=a_2\}.
\]
Every initial, transition, and observation probability is positive.

For each joint action \(a\), define the expected successor-state reward
component and the marginalized next-observation distribution by
\[
m_a=\sum_{s'}p_a(s')c(s'),
\qquad
\bar q_a(o')=\sum_{s'}p_a(s')q_{s'}(o').
\]
For example,
\[
\begin{aligned}
m_{01}&=\tfrac13c(L)+\tfrac13c(M)+\tfrac13c(H)=1,
\\
\bar q_{01}&=\tfrac13(q_L+q_M+q_H)=(1/4,1/4,1/4,1/4).
\end{aligned}
\]
The same calculations for the other joint actions give
\[
m_{00}=\frac23,
\qquad
m_{01}=m_{10}=1,
\qquad
m_{11}=\frac43,
\]
and
\[
\begin{aligned}
\bar q_{00}&=(7/24,1/4,1/4,5/24),\\
\bar q_{01}&=\bar q_{10}=(1/4,1/4,1/4,1/4),\\
\bar q_{11}&=(5/24,1/4,1/4,7/24).
\end{aligned}
\]
Writing \(u_i(a)=2a_i+\mathbf 1\{a_1=a_2\}\), the transformation \(g\),
whose agent permutation swaps the two agents, therefore satisfies
\[
m_{g\cdot a}=m_a,
\qquad
\bar q_{g\cdot a}(g\cdot o')=\bar q_a(o'),
\qquad
u_{\sigma_g(i)}(g\cdot a)=u_i(a).
\]
At \(t=1\),
we separate the current-state contribution \(c(s)\) from the remaining
expected reward by defining an expression, such that:
\[
W_i^\pi(h^1)
=\sum_{a\in\mathcal A}\pi^1(a\mid h^1)[u_i(a)+m_a].
\]
The Bellman recursion at \(t=1\) reduces to
\[
V_{i,1}^\pi(s,h^1)=c(s)+W_i^\pi(h^1).
\]
The policy transformation and the three displayed identities give
\[
W_{\sigma_g(i)}^{g\cdot\pi}(g\cdot h^1)=W_i^\pi(h^1).
\]
Indeed, changing variables by \(b=g\cdot a\) in the left-hand side matches
the policy probability, \(u_{\sigma_g(i)}(b)\), and \(m_b\) with their three
source terms on the right-hand side.
At the initial time step \(t=0\), expanding Equation~\eqref{eq:bellman} yields the following equation:
\[
\begin{aligned}
V_{i,0}^{\pi}(s,h^0)
=\sum_{a\in\mathcal A}\pi^0(a\mid h^0)
\Bigl[&c(s)+u_i(a)+(1+\gamma)m_a\\
&+\gamma\sum_{o'\in\mathcal O}\bar q_a(o')
W_i^\pi(h^0,a,o')\Bigr].
\end{aligned}
\]
The coefficient \(1+\gamma\) appears because \(c(s')\) contributes once to
the immediate reward and once, with discount \(\gamma\), to the value at
\(t=1\).  In the transformed expression, use the bijections
\(b=g\cdot a\) and \(\bar o'=g\cdot o'\).  The policy transformation matches the
action probability.  The identities for \(u_i\) and \(m_a\) preserve the
terms \(u_i(a)\) and \((1+\gamma)m_a\), respectively.  The
identities for \(\bar q_a\) and \(W_i^\pi\) preserve the remaining term
\(\gamma\sum_{o'\in\mathcal O}\bar q_a(o')W_i^\pi(h^0,a,o')\).
Consequently, every term of the transformed sum equals its
corresponding source term, which proves the value identity at the initial time step.
Together with
the identity at \(t=1\) and the zero values at \(t=2\), this gives that:
\[
V_{\sigma_g(i),t}^{g\cdot\pi}(\tau_g^t x_t)
=V_{i,t}^{\pi}(x_t)
\]
for every policy, agent, time step, and valid state--joint-history pair.  Thus
\(g\in\Geval\).  Its state map is the identity, so it also preserves the
initial distribution.

Nevertheless, for \(a=(0,1)\) and \(s'=M\),
\[
P(M\mid s,g\cdot a)=\frac16
\neq\frac13=P(M\mid s,a),
\]
so \(g\notin\AutEnv(\mathcal M)\).  It remains to rule out a different
componentwise environment automorphism that exchanges the agents.

Suppose, for a contradiction, that there exists a componentwise environment
automorphism \(\widetilde g\) that exchanges the two agents.  Let \(\psi\)
denote its state permutation, and allow arbitrary binary action and
observation relabelings.  Substitute
\(r_i(s,a,s')=c(s)+c(s')+u_i(a)\) into the reward-preservation equation and
move the source terms to the left.  This gives the following equations:
\[
d(s)+d(s')
+u_{\sigma_{\widetilde g}(i)}(\widetilde g\cdot a)-u_i(a)=0,
\qquad
d(s)=c(\psi(s))-c(s).
\]
Holding \(s',a,i\) fixed shows that \(d(s)\) is constant.  Because \(\psi\)
only permutes the three states,
\[
\sum_s d(s)=\sum_sc(\psi(s))-\sum_sc(s)=0.
\]
A constant function with zero sum is zero.  Hence, \(d(s)=0\) for every state,
and injectivity of \(c\) gives
\(\psi=\operatorname{id}_{\mathcal S}\).  Reward preservation reduces to an equation, such that:
\[
u_{\sigma_{\widetilde g}(i)}(\widetilde g\cdot a)=u_i(a).
\]
Write the two local action-label maps as \(f_1\) and \(f_2\).  A bijection of
the binary action set is either the identity or the label flip.  Under the
agent swap,
\([\widetilde g\cdot a]_2=f_1(a_1)\) and
\([\widetilde g\cdot a]_1=f_2(a_2)\).  If \(f_1\) flips labels, choose source
agent \(1\) and \(a_1=0\).  Then,
\[
u_2(\widetilde g\cdot a)
=2f_1(0)+\mathbf 1\{f_2(a_2)=f_1(0)\}\geq2,
\]
whereas \(u_1(a)\leq1\), a contradiction.  Thus \(f_1\) is the identity.
Repeating the same argument for source agent \(2\) shows that \(f_2\) is also
the identity.  Observation relabeling
cannot repair the transition mismatch because the transition equation does
not contain an observation map.  Therefore no componentwise environment
automorphism exchanges the agents, while \(g\in\Geval\) does.  Together with
Theorem~\ref{thm:componentwise-certificate}, which gives
\(\AutEnv(\mathcal M)\leq\Geval\), this proves
\(\AutEnv(\mathcal M)\lneq\Geval\).  The two orbit partitions are
consequently
\[
\mathcal P(\AutEnv(\mathcal M))=\{\{1\},\{2\}\},
\qquad
\Peval(\mathcal M)=\{\{1,2\}\}.
\]
All transition, observation, and initial-state probabilities are positive,
so the separation is not an artifact of histories that no policy can reach.
\end{proof}

\subsection{Subgroup, orbit, and normal-form facts}

\begin{proposition}[Policy-evaluation symmetry subgroup and return equivariance
under the initial distribution]
\label{prop:supp-policy-evaluation-subgroup}
The set \(\Geval\) is a subgroup of \(\RAdm\).  Let
\[
\Gevali
=\left\{g\in\Geval:
\rho_0\bigl(\phi_g^{\mathcal S}(s)\bigr)=\rho_0(s)
\text{ for every }s\in\mathcal S\right\}.
\]
Then \(\Gevali\) is a subgroup, and every \(g\in\Gevali\) satisfies
\[
J_{\sigma_g(i)}(g\cdot\pi)=J_i(\pi)
\qquad(i\in N,\ \pi\in\Pi_{\mathrm{dec}}).
\]
\end{proposition}

\begin{proof}
We verify the three subgroup properties directly.  The identity transformation
satisfies the evaluation identity.  If \(g,k\in\Geval\), then
\[
\begin{aligned}
\mathcal E_{\mathcal M}((gk)\cdot\pi)
&=\mathcal E_{\mathcal M}(g\cdot(k\cdot\pi))\\
&=g\cdot\mathcal E_{\mathcal M}(k\cdot\pi)
=(gk)\cdot\mathcal E_{\mathcal M}(\pi).
\end{aligned}
\]
For inverses, fix \(g\in\Geval\) and any \(\pi\in\Pi_{\mathrm{dec}}\).
Since \(g^{-1}\in\RAdm\), Proposition~\ref{prop:supp-induced-actions}
ensures that \(g^{-1}\cdot\pi\in\Pi_{\mathrm{dec}}\).  Applying the defining
evaluation identity for \(g\) to this policy and using
\(g\cdot(g^{-1}\cdot\pi)=\pi\) gives the following equation:
\[
\mathcal E_{\mathcal M}(\pi)
=g\cdot\mathcal E_{\mathcal M}(g^{-1}\cdot\pi).
\]
Apply \(g^{-1}\) to both value profiles.  The value-profile action law
\(g^{-1}\cdot(g\cdot V)=V\) then gives the following equation:
\[
\mathcal E_{\mathcal M}(g^{-1}\cdot\pi)
=g^{-1}\cdot\mathcal E_{\mathcal M}(\pi).
\]
Since \(\pi\) was arbitrary, this is the defining evaluation identity for
\(g^{-1}\), so \(g^{-1}\in\Geval\).

The additional \(\rho_0\)-preservation condition is also closed under the
group operations.  The identity state map preserves \(\rho_0\).  If \(g\)
and \(k\) preserve it, then we have that:
\[
\rho_0\bigl(\phi_{gk}^{\mathcal S}(s)\bigr)
=\rho_0\bigl(\phi_g^{\mathcal S}(\phi_k^{\mathcal S}(s))\bigr)
=\rho_0(\phi_k^{\mathcal S}(s))
=\rho_0(s).
\]
If \(g\) preserves \(\rho_0\), substituting
\(u=(\phi_g^{\mathcal S})^{-1}(s)\) into
\(\rho_0(\phi_g^{\mathcal S}(u))=\rho_0(u)\) shows that \(g^{-1}\) also
preserves it.  Hence \(\Gevali\) is a subgroup.

Finally, let \(g\in\Gevali\).  Preservation of the initial distribution
means
\(\rho_0(\phi_g^{\mathcal S}(s))=\rho_0(s)\).  To apply evaluation
equivariance, we reindex the initial-state sum using
\(s'=\phi_g^{\mathcal S}(s)\).  Bijectivity ensures that every state is
counted exactly once, and initial-distribution preservation ensures that
its probability weight is unchanged.  The empty initial joint history is
fixed by the transformation, so we obtain the following identity:
\[
\begin{aligned}
J_{\sigma_g(i)}(g\cdot\pi)
&=\sum_{s'\in\mathcal S}
  \rho_0(s')V_{\sigma_g(i),0}^{g\cdot\pi}(s',()^N)\\
&=\sum_{s\in\mathcal S}
  \rho_0(\phi_g^{\mathcal S}(s))
  V_{\sigma_g(i),0}^{g\cdot\pi}(\phi_g^{\mathcal S}(s),()^N)\\
&=\sum_{s\in\mathcal S}\rho_0(s)V_{i,0}^{\pi}(s,()^N)
=J_i(\pi).
\end{aligned}
\]
\end{proof}

\begin{proposition}[Orbit facts]
\label{prop:supp-orbit-facts}
For every subgroup
\[
H\leq\RAdm,
\]
define
\(i\sim_H j\) when some \(g\in H\) satisfies \(\sigma_g(i)=j\).
Then \(\sim_H\) is an equivalence relation.  If \(H\leq K\), then
\(\mathcal P(H)\preceq\mathcal P(K)\).
\end{proposition}

\begin{proof}
We first show that \(\sim_H\) is an equivalence relation by verifying
reflexivity, symmetry, and transitivity~\citep[Chap.~1.2]{judson1994abstract}.

For reflexivity, we must show that \(i\sim_H i\) for every agent \(i\).
Because \(H\) is a subgroup, it contains the identity transformation \(e\).
This transformation leaves every agent label unchanged, so \(\sigma_e(i)=i\).
Thus, \(e\in H\) satisfies the defining condition for \(i\sim_H i\).

For symmetry, suppose that \(i\sim_H j\).  By definition, some \(g\in H\)
satisfies \(\sigma_g(i)=j\).  Since \(H\) is a subgroup, the inverse
transformation \(g^{-1}\) also belongs to \(H\).  Its agent permutation reverses
that of \(g\), giving
\[
\begin{aligned}
\sigma_{g^{-1}}(j) =\sigma_{g^{-1}}\bigl(\sigma_g(i)\bigr) =\sigma_{g^{-1}g}(i)=\sigma_e(i)=i.
\end{aligned}
\]
Thus, \(g^{-1}\in H\) maps agent \(j\) back to agent \(i\), which establishes
\(j\sim_H i\).  This proves symmetry of the relation.

For transitivity, if \(\sigma_g(i)=j\) and \(\sigma_k(j)=\ell\) for
\(g,k\in H\), then \(kg\in H\) and \(\sigma_{kg}(i)=\ell\), giving
\(i\sim_H\ell\).  The equivalence classes are
precisely the \(H\)-orbits, which form the partition \(\mathcal P(H)\).

If \(H\leq K\), then we have
\[
\{\sigma_g(i):g\in H\}
\subseteq
\{\sigma_g(i):g\in K\},
\]
for every agent \(i\).  Thus, every \(H\)-orbit is contained in a \(K\)-orbit.
Since these orbits are the parts of the corresponding partitions, every
part of \(\mathcal P(H)\) is contained in a part of \(\mathcal P(K)\).
By the definition of partition refinement,
\(\mathcal P(H)\preceq\mathcal P(K)\).
\end{proof}

To check the definition in a static setting, we show that policy-evaluation
symmetry reduces to payoff preservation for the following one-stage
normal-form embedding.

\begin{proposition}[Normal-form special case]
\label{prop:supp-normal-form}
Embed a finite normal-form game
\(\Gamma=(N,(\mathcal A_i)_i,(u_i)_i)\) as a horizon-one POSG with one
state, singleton observations, deterministic dynamics, and reward
\(r_i(s_\star,a,s_\star)=u_i(a)\).  For an admissible transformation \(g\),
evaluation equivariance for every mixed product policy is equivalent to
\[
u_{\sigma_g(i)}(g\cdot a)=u_i(a)
\qquad\text{for every }i\text{ and pure profile }a.
\]
Every admissible transformation automatically preserves the unique initial
state--joint-history distribution.
\end{proposition}

\begin{proof}
Let \(U_i(\pi)\) denote agent \(i\)'s expected payoff under the mixed product
policy \(\pi\).  If the displayed payoff identity holds, reindexing joint
actions and using the product-policy transformation gives
\[
\begin{aligned}
U_{\sigma_g(i)}(g\cdot\pi)
&=\sum_{a\in\mathcal A}
u_{\sigma_g(i)}(g\cdot a)\prod_{j\in N}\pi_j(a_j)\\
&=\sum_{a\in\mathcal A}u_i(a)\prod_{j\in N}\pi_j(a_j)
=U_i(\pi).
\end{aligned}
\]
Conversely, the policy domain contains every deterministic product policy
\(\delta_a\), and \(g\cdot\delta_a=\delta_{g\cdot a}\).  Evaluation
equivariance gives
\[
u_{\sigma_g(i)}(g\cdot a)
=U_{\sigma_g(i)}(\delta_{g\cdot a})
=U_i(\delta_a)=u_i(a).
\]
Finally, let \(g\in\RAdm\) be arbitrary.  Since
\(\mathcal S=\{s_\star\}\), its bijective state map must satisfy
\(\phi_g^{\mathcal S}(s_\star)=s_\star\).  Every initial local history is
empty, so there are no actions or observations to transform.  Permuting the
agents only rearranges identical empty histories, giving
\(g\cdot()^N=()^N\).  Consequently,
\[
\begin{aligned}
\tau_g^0(s_\star,()^N)
&=\bigl(\phi_g^{\mathcal S}(s_\star),g\cdot()^N\bigr)\\
&=(s_\star,()^N).
\end{aligned}
\]
The initial distribution assigns probability one to this single pair.
Because the pair is fixed, its distribution is preserved by every
admissible transformation.
\end{proof}

\section{Proofs for Componentwise Environment Automorphisms}
\label{app:componentwise-proofs}

\begin{proposition}[Subgroup property of componentwise environment automorphisms]
\label{prop:supp-componentwise-subgroup}
The set in Definition~\ref{def:componentwise} satisfies
\[
\AutEnv(\mathcal M)\leq\RAdm.
\]
The componentwise environment automorphisms that preserve the initial
distribution, such that:
\[
\left\{g\in\AutEnv(\mathcal M):
\rho_0\bigl(\phi_g^{\mathcal S}(s)\bigr)=\rho_0(s)
\text{ for every }s\in\mathcal S\right\},
\]
also form a subgroup.
\end{proposition}

\begin{proof}
The identity preserves each component.  Suppose \(g\) and \(k\) satisfy
Equations~\eqref{eq:component-P}--\eqref{eq:component-r}.  For transitions,
\[
\begin{aligned}
&P(\phi_{gk}^{\mathcal S}(s')\mid
\phi_{gk}^{\mathcal S}(s),(gk)\cdot a)\\
&=P(\phi_g^{\mathcal S}(\phi_k^{\mathcal S}(s'))\mid
\phi_g^{\mathcal S}(\phi_k^{\mathcal S}(s)),g\cdot(k\cdot a))\\
&=P(\phi_k^{\mathcal S}(s')\mid
\phi_k^{\mathcal S}(s),k\cdot a)
=P(s'\mid s,a).
\end{aligned}
\]
The first preservation equality removes the outer transformation \(g\), and
the second removes \(k\).  The observation and reward calculations have the
same two-step procedure.  For observations,
\[
\begin{aligned}
&Z((gk)\cdot o'\mid\phi_{gk}^{\mathcal S}(s'),(gk)\cdot a)\\
&=Z(g\cdot(k\cdot o')\mid
\phi_g^{\mathcal S}(\phi_k^{\mathcal S}(s')),g\cdot(k\cdot a))\\
&=Z(k\cdot o'\mid\phi_k^{\mathcal S}(s'),k\cdot a)
=Z(o'\mid s',a).
\end{aligned}
\]
For rewards, tracking the agent coordinate gives
\[
\begin{aligned}
&r_{\sigma_{gk}(i)}(\phi_{gk}^{\mathcal S}(s),(gk)\cdot a,
\phi_{gk}^{\mathcal S}(s'))\\
&=r_{\sigma_g(\sigma_k(i))}
(\phi_g^{\mathcal S}(\phi_k^{\mathcal S}(s)),g\cdot(k\cdot a),
\phi_g^{\mathcal S}(\phi_k^{\mathcal S}(s')))\\
&=r_{\sigma_k(i)}(\phi_k^{\mathcal S}(s),k\cdot a,
\phi_k^{\mathcal S}(s'))
=r_i(s,a,s').
\end{aligned}
\]
Thus the set is closed under composition.

For inverse closure, fix \(g\in\AutEnv(\mathcal M)\).  Since \(\RAdm\) is
a group, \(g^{-1}\) is an admissible transformation.  We show that it also
preserves the three model components by applying the identities already
known for \(g\) to suitable inputs.

For transitions, Equation~\eqref{eq:component-P} holds for every pair of
states \(u,u'\) and joint action \(b\):
\[
P(\phi_g^{\mathcal S}(u')\mid\phi_g^{\mathcal S}(u),g\cdot b)
=P(u'\mid u,b).
\]
For arbitrary \(s,s'\), and \(a\), choose
\[
u=\phi_{g^{-1}}^{\mathcal S}(s),\qquad
u'=\phi_{g^{-1}}^{\mathcal S}(s'),\qquad
b=g^{-1}\cdot a.
\]
Applying \(g\) cancels these inverse transformations, so
\(\phi_g^{\mathcal S}(u)=s\), \(\phi_g^{\mathcal S}(u')=s'\), and
\(g\cdot b=a\).  Substitution into the transition identity therefore gives
\[
P(s'\mid s,a)
=P(\phi_{g^{-1}}^{\mathcal S}(s')\mid
\phi_{g^{-1}}^{\mathcal S}(s),g^{-1}\cdot a),
\]
which shows that the transition probabilities are preserved under
\(g^{-1}\).

For observations, Equation~\eqref{eq:component-Z} gives
\[
Z(g\cdot v'\mid\phi_g^{\mathcal S}(u'),g\cdot b)
=Z(v'\mid u',b).
\]
Use the same \(u'\) and \(b\), and set \(v'=g^{-1}\cdot o'\) for an
arbitrary joint observation \(o'\).  Since \(g\cdot v'=o'\), substitution
gives
\[
Z(o'\mid s',a)
=Z(g^{-1}\cdot o'\mid
\phi_{g^{-1}}^{\mathcal S}(s'),g^{-1}\cdot a).
\]

For rewards, Equation~\eqref{eq:component-r} holds for every agent \(j\):
\[
r_{\sigma_g(j)}(\phi_g^{\mathcal S}(u),g\cdot b,
\phi_g^{\mathcal S}(u'))=r_j(u,b,u').
\]
Use the same \(u,u'\), and \(b\), and choose
\(j=\sigma_{g^{-1}}(i)\) for an arbitrary agent \(i\).  The agent
permutations also cancel, giving \(\sigma_g(j)=i\).  Thus, substitution
into the reward identity gives
\[
r_i(s,a,s')
=r_{\sigma_{g^{-1}}(i)}(\phi_{g^{-1}}^{\mathcal S}(s),g^{-1}\cdot a,
\phi_{g^{-1}}^{\mathcal S}(s')).
\]
Because \(s,s',a,o'\), and \(i\) were arbitrary, these three identities
show that \(g^{-1}\) preserves the transition probabilities, observation
probabilities, and reward functions.  Hence,
\(g^{-1}\in\AutEnv(\mathcal M)\).  Together with the identity and composition
properties proved above, this establishes that \(\AutEnv(\mathcal M)\) is a
subgroup.

It remains to impose preservation of the initial distribution.  The identity
state map preserves \(\rho_0\).  If \(g\) and \(k\) preserve \(\rho_0\), then
\[
\rho_0\bigl(\phi_{gk}^{\mathcal S}(s)\bigr)
=\rho_0\bigl(\phi_g^{\mathcal S}(\phi_k^{\mathcal S}(s))\bigr)
=\rho_0\bigl(\phi_k^{\mathcal S}(s)\bigr)
=\rho_0(s).
\]
For inverse closure, suppose \(g\) preserves \(\rho_0\).  This means that
for every state \(u\in\mathcal S\),
\[
\rho_0\bigl(\phi_g^{\mathcal S}(u)\bigr)=\rho_0(u).
\]
For an arbitrary state \(s\), choose
\(u=\phi_{g^{-1}}^{\mathcal S}(s)=(\phi_g^{\mathcal S})^{-1}(s)\).
Substituting this state into the preservation identity for \(g\) gives
\[
\rho_0\bigl(\phi_g^{\mathcal S}(\phi_{g^{-1}}^{\mathcal S}(s))\bigr)
=\rho_0\bigl(\phi_{g^{-1}}^{\mathcal S}(s)\bigr).
\]
Since \(\phi_g^{\mathcal S}(\phi_{g^{-1}}^{\mathcal S}(s))=s\), the
left-hand side simplifies to \(\rho_0(s)\), yielding
\[
\rho_0(s)=\rho_0\bigl(\phi_{g^{-1}}^{\mathcal S}(s)\bigr).
\]
This holds for every \(s\), so \(g^{-1}\) also preserves \(\rho_0\).
Together with the componentwise closure already proved, these calculations
establish the second subgroup claim.
\end{proof}

\begin{proof}[Proof of Theorem~\ref{thm:componentwise-certificate}]
Proposition~\ref{prop:supp-componentwise-subgroup} already establishes that
\(\AutEnv(\mathcal M)\) is a subgroup of \(\RAdm\).  We now show that
every element of this subgroup is a policy-evaluation symmetry.
Fix \(g\in\AutEnv(\mathcal M)\).  We prove by backward induction that
\[
V_{\sigma_g(i),t}^{g\cdot\pi}(\tau_g^t x_t)
=V_{i,t}^{\pi}(x_t)
\]
for every agent \(i\), policy \(\pi\), time step \(t\), and
\(x_t\in\mathcal X_t\).  At \(t=T\), both corresponding values are zero.
Assume the identity holds at time step \(t+1\).  For \(t<T\), expand the
transformed value:
\[
\begin{aligned}
&V_{\sigma_g(i),t}^{g\cdot\pi}(\tau_g^t(s,h^t))\\
&=\sum_{b,\bar s',\bar o'}
(g\cdot\pi)^t(b\mid g\cdot h^t)
P(\bar s'\mid\phi_g^{\mathcal S}(s),b)
Z(\bar o'\mid\bar s',b)\\
&\quad\cdot\Bigl[
r_{\sigma_g(i)}(\phi_g^{\mathcal S}(s),b,\bar s')
+\gamma V_{\sigma_g(i),t+1}^{g\cdot\pi}
(\bar s',g\cdot h^t,b,\bar o')
\Bigr].
\end{aligned}
\]
Reindex the sum using
\[
b=g\cdot a,
\qquad
\bar s'=\phi_g^{\mathcal S}(s'),
\qquad
\bar o'=g\cdot o'.
\]
Each transformation is bijective, so every triple \((b,\bar s',\bar o')\)
corresponds to exactly one triple \((a,s',o')\).

For the joint-action probability, the product-policy definition and the
local policy-transformation rule in Equation~\eqref{eq:policy-action} give
\[
\begin{aligned}
&(g\cdot\pi)^t(g\cdot a\mid g\cdot h^t)\\
&=\prod_{j\in N}(g\cdot\pi)_{\sigma_g(j)}^t
\bigl(\phi_{g,j}^{\mathcal A}(a_j)\mid
\Phi_{g,j}^{\mathcal H,t}(h_j^t)\bigr)\\
&=\prod_{j\in N}\pi_j^t(a_j\mid h_j^t)
=\pi^t(a\mid h^t).
\end{aligned}
\]
The first equality indexes the product by \(\sigma_g(j)\), which visits
each agent exactly once because \(\sigma_g\) is a permutation.

Since \(g\in\AutEnv(\mathcal M)\), Definition~\ref{def:componentwise}
provides Equations~\eqref{eq:component-P}--\eqref{eq:component-r}.
Applying these identities to the substituted transition, observation, and
reward terms gives
\[
\begin{aligned}
P(\phi_g^{\mathcal S}(s')\mid\phi_g^{\mathcal S}(s),g\cdot a)
&=P(s'\mid s,a),\\
Z(g\cdot o'\mid\phi_g^{\mathcal S}(s'),g\cdot a)
&=Z(o'\mid s',a),\\
r_{\sigma_g(i)}(\phi_g^{\mathcal S}(s),g\cdot a,
\phi_g^{\mathcal S}(s'))
&=r_i(s,a,s').
\end{aligned}
\]

To apply the induction hypothesis to the remaining value term, for each
joint action \(a\) and next joint observation \(o'\), write
\[
h^{t+1}:=(h^t,a,o').
\]
This is the joint history of length \(t+1\) obtained by appending \(a\)
and \(o'\) to \(h^t\).
The history transformation applies the same action and observation maps to
every entry and preserves their order.  Thus, transforming this extended
history gives
\[
g\cdot h^{t+1}=(g\cdot h^t,g\cdot a,g\cdot o').
\]
Recall from Section~\ref{sec:setting} that
\[
\tau_g^t(s,h^t)=\bigl(\phi_g^{\mathcal S}(s),g\cdot h^t\bigr),
\]
so \(\tau_g^t\) transforms both the state and the joint history at time step
\(t\).  Applying this definition at time step \(t+1\) and using the
preceding history identity, the next state--joint-history pair in the
substituted sum is therefore
\[
\begin{aligned}
&\bigl(\phi_g^{\mathcal S}(s'),(g\cdot h^t,g\cdot a,g\cdot o')\bigr)\\
&=\bigl(\phi_g^{\mathcal S}(s'),g\cdot h^{t+1}\bigr)\\
&=\tau_g^{t+1}\bigl(s',h^{t+1}\bigr).
\end{aligned}
\]
The induction hypothesis at time step \(t+1\), applied to
\((s',h^{t+1})\), now gives
\[
\begin{aligned}
&V_{\sigma_g(i),t+1}^{g\cdot\pi}
\bigl(\phi_g^{\mathcal S}(s'),g\cdot h^{t+1}\bigr)\\
&=V_{\sigma_g(i),t+1}^{g\cdot\pi}
\bigl(\tau_g^{t+1}(s',h^{t+1})\bigr)\\
&=V_{i,t+1}^{\pi}\bigl(s',h^{t+1}\bigr).
\end{aligned}
\]
Using the preceding identities in sequence, the Bellman expansion becomes
\[
\begin{aligned}
&V_{\sigma_g(i),t}^{g\cdot\pi}(\tau_g^t(s,h^t))\\
&\overset{(1)}{=}\sum_{a,s',o'}
(g\cdot\pi)^t(g\cdot a\mid g\cdot h^t)\\
&\quad\cdot P(\phi_g^{\mathcal S}(s')\mid
\phi_g^{\mathcal S}(s),g\cdot a)\\
&\quad\cdot Z(g\cdot o'\mid\phi_g^{\mathcal S}(s'),g\cdot a)\\
&\quad\cdot\Bigl[
r_{\sigma_g(i)}(\phi_g^{\mathcal S}(s),g\cdot a,
\phi_g^{\mathcal S}(s'))\\
&\qquad\quad+\gamma V_{\sigma_g(i),t+1}^{g\cdot\pi}
\bigl(\tau_g^{t+1}(s',h^{t+1})\bigr)\Bigr]\\[2pt]
&\overset{(2)}{=}\sum_{a,s',o'}
\pi^t(a\mid h^t)P(s'\mid s,a)Z(o'\mid s',a)\\
&\quad\cdot\Bigl[r_i(s,a,s')
+\gamma V_{\sigma_g(i),t+1}^{g\cdot\pi}
\bigl(\tau_g^{t+1}(s',h^{t+1})\bigr)\Bigr]\\[2pt]
&\overset{(3)}{=}\sum_{a,s',o'}
\pi^t(a\mid h^t)P(s'\mid s,a)Z(o'\mid s',a)\\
&\quad\cdot\Bigl[r_i(s,a,s')
+\gamma V_{i,t+1}^{\pi}\bigl(s',h^{t+1}\bigr)\Bigr]\\
&\overset{(4)}{=}V_{i,t}^{\pi}(s,h^t).
\end{aligned}
\]
Step~(1) uses the bijective substitutions and the identity for the next
state--joint-history pair.  Step~(2) uses the joint-policy identity derived
from Equation~\eqref{eq:policy-action} and the componentwise identities in
Equations~\eqref{eq:component-P}--\eqref{eq:component-r} to replace the
action, transition, and observation probabilities and the reward.
Step~(3) applies the induction hypothesis at time step \(t+1\) to the
remaining transformed value.  Step~(4) is Equation~\eqref{eq:bellman}, with
\(h^{t+1}=(h^t,a,o')\).  This completes the induction.
The equality holds for every agent, policy, time step,
and valid state--joint-history pair, so Definition~\ref{def:policy-evaluation}
gives \(g\in\Geval\).

Since \(g\) was arbitrary, \(\AutEnv(\mathcal M)\subseteq\Geval\).
The subgroup property from
Proposition~\ref{prop:supp-componentwise-subgroup} then gives
\(\AutEnv(\mathcal M)\leq\Geval\).
Proposition~\ref{prop:supp-orbit-facts} yields the corresponding partition
refinement:
\[
\mathcal P(\AutEnv(\mathcal M))\preceq\Peval(\mathcal M),
\]
as claimed.
\end{proof}

\section{Extension to Dec-POMDPs}
\label{app:cooperative-proofs}

\begin{proposition}[Existing common reward]
\label{prop:supp-common-reward}
If \(r_i=r^{\mathrm{com}}\) for every agent, then, for every policy \(\pi\),
time step \(t\), valid \(x\in\mathcal X_t\), and agents \(i,j\),
\[
V_{i,t}^{\pi}(x)=V_{j,t}^{\pi}(x).
\]
The structural agent types nevertheless remain the orbits of
\(G_{\mathcal M}^{\mathrm{eval}}\); in particular, equality of the value coordinates
does not by itself imply that this group acts transitively on the agents.
\end{proposition}

\begin{proof}
Fix an arbitrary policy \(\pi\in\Pi_{\mathrm{dec}}\) and agents \(i,j\in N\).
We prove equality of their values at every valid input by backward induction.
At \(t=T\),
\[
V_{i,T}^{\pi}(x)=0=V_{j,T}^{\pi}(x)
\qquad\text{for every }x\in\mathcal X_T.
\]
Suppose that \(V_{i,t+1}^{\pi}(x')=V_{j,t+1}^{\pi}(x')\) for every
\(x'\in\mathcal X_{t+1}\).  Fix any \((s,h^t)\in\mathcal X_t\), and
write \(h^{t+1}=(h^t,a,o')\) in each summand.  Equation~\eqref{eq:bellman}
and \(r_i=r_j=r^{\mathrm{com}}\) give
\[
\begin{aligned}
V_{i,t}^{\pi}(s,h^t)
&=\sum_{a,s',o'}\pi^t(a\mid h^t)P(s'\mid s,a)Z(o'\mid s',a)\\
&\quad\cdot\Bigl[r^{\mathrm{com}}(s,a,s')
+\gamma V_{i,t+1}^{\pi}(s',h^{t+1})\Bigr]\\
&=\sum_{a,s',o'}\pi^t(a\mid h^t)P(s'\mid s,a)Z(o'\mid s',a)\\
&\quad\cdot\Bigl[r^{\mathrm{com}}(s,a,s')
+\gamma V_{j,t+1}^{\pi}(s',h^{t+1})\Bigr]\\
&=V_{j,t}^{\pi}(s,h^t).
\end{aligned}
\]
The middle equality uses the induction hypothesis at the next input
\((s',h^{t+1})\), which lies in \(\mathcal X_{t+1}\).
The first and last equalities use
the Bellman recursion with the common reward.  This proves the induction
step.  Since \(\pi,i,j\), and the input were arbitrary, the value equality
holds with all the stated quantifiers.

The orbit characterization of structural agent types follows directly from
Definition~\ref{def:structural-type}.  To show that the value equality does
not imply transitivity, consider the one-stage common-reward example from
Section~\ref{sec:cooperative}.  Specify
\(N=\{1,2\}\), \(T=1\), \(\mathcal S=\{s_\star\}\),
\(\mathcal A_1=\mathcal A_2=\{0,1\}\), and
\(\mathcal O_1=\mathcal O_2=\{o_\star\}\).
The initial state is \(s_\star\) with probability one, and the transition
and observation distributions assign probability one to the unique next
state and joint observation.  Both agents receive
\[
r_1(s_\star,a,s_\star)=r_2(s_\star,a,s_\star)=a_1.
\]
Take \(\RAdm\) to be the full group of type-compatible transformations, so
agent exchanges and arbitrary binary action relabelings are admissible.

Suppose, for a contradiction, that some \(g\in\Geval\) exchanges the
two agents.  The first coordinate of \(g\cdot a\) is then
\(\phi_{g,2}^{\mathcal A}(a_2)\).  By
Proposition~\ref{prop:supp-normal-form}, \(g\) must preserve the payoffs
at every pure joint action.  The common reward therefore requires
\[
\phi_{g,2}^{\mathcal A}(a_2)=a_1
\qquad\text{for every }a_1,a_2\in\{0,1\}.
\]
Taking \((a_1,a_2)=(0,0)\) and \((1,0)\) would give both
\(\phi_{g,2}^{\mathcal A}(0)=0\) and
\(\phi_{g,2}^{\mathcal A}(0)=1\), a contradiction.  Hence no
policy-evaluation symmetry exchanges the two agents, and
\[
\Peval(\mathcal M)=\bigl\{\{1\},\{2\}\bigr\}.
\]
The value coordinates are equal by the first part of this proof, but the
policy-evaluation symmetry group is not transitive.  This proves the final
claim.
\end{proof}

\begin{proof}[Proof of Theorem~\ref{thm:weighted-aggregation}]
Fix an arbitrary policy \(\pi\in\Pi_{\mathrm{dec}}\).  We first prove the
weighted-value identity for every derived agent \(j\) and every valid input
by backward induction on the time step.

At \(t=T\), all values in both models are zero.
Consequently, for every \(j\in N\) and \(x\in\mathcal X_T\),
\[
V_{j,T}^{\pi;\mathcal M_w}(x)
=0=\sum_{i\in N}w_i\,0
=\sum_{i\in N}w_iV_{i,T}^{\pi;\mathcal M}(x).
\]
This establishes the base case.

For the induction step, let \(t<T\) and suppose that
\[
V_{j,t+1}^{\pi;\mathcal M_w}(x')
=\sum_{i\in N}w_iV_{i,t+1}^{\pi;\mathcal M}(x')
\]
for every \(j\in N\) and every \(x'\in\mathcal X_{t+1}\).
We must establish the same identity at time step \(t\).  Fix any
\(j\in N\) and \(x=(s,h^t)\in\mathcal X_t\).  For each action \(a\),
next state \(s'\), and next observation \(o'\) in the Bellman sum, write
\[
h^{t+1}:=(h^t,a,o'),
\qquad x':=(s',h^{t+1})\in\mathcal X_{t+1}.
\]
Thus the induction hypothesis applies to every next input in that sum.
The two models have the same policy domain, transition and observation
distributions, and discount factor.  Evaluating the fixed policy \(\pi\)
therefore uses the same factors \(\pi^t(a\mid h^t)\),
\(P(s'\mid s,a)\), and \(Z(o'\mid s',a)\) in both models.
The Bellman recursion gives the following sequence of equalities:
\[
\begin{aligned}
&V_{j,t}^{\pi;\mathcal M_w}(s,h^t)\\
&\overset{(1)}{=}\sum_{a,s',o'}\pi^t(a\mid h^t)P(s'\mid s,a)Z(o'\mid s',a)\\
&\quad\cdot\Bigl[r^w(s,a,s')
+\gamma V_{j,t+1}^{\pi;\mathcal M_w}(x')\Bigr]\\
&\overset{(2)}{=}\sum_{a,s',o'}\pi^t(a\mid h^t)P(s'\mid s,a)Z(o'\mid s',a)\\
&\quad\cdot\Biggl[\sum_{i\in N}w_i r_i(s,a,s')
+\gamma\sum_{i\in N}w_iV_{i,t+1}^{\pi;\mathcal M}(x')\Biggr]\\
&\overset{(3)}{=}\sum_{a,s',o'}\pi^t(a\mid h^t)P(s'\mid s,a)Z(o'\mid s',a)\\
&\quad\cdot\sum_{i\in N}w_i
 \Bigl[r_i(s,a,s')+\gamma V_{i,t+1}^{\pi;\mathcal M}(x')\Bigr]\\
&\overset{(4)}{=}\sum_{i\in N}w_i\sum_{a,s',o'}
 \pi^t(a\mid h^t)P(s'\mid s,a)Z(o'\mid s',a)\\
&\quad\cdot\Bigl[r_i(s,a,s')
+\gamma V_{i,t+1}^{\pi;\mathcal M}(x')\Bigr]\\
&\overset{(5)}{=}\sum_{i\in N}w_iV_{i,t}^{\pi;\mathcal M}(s,h^t).
\end{aligned}
\]
Step (1) is Equation~\eqref{eq:bellman} in \(\mathcal M_w\), where agent
\(j\)'s reward is \(r^w\).  Step (2) replaces \(r^w\) by its defining
weighted sum and applies the induction hypothesis to
\(V_{j,t+1}^{\pi;\mathcal M_w}(x')\).  Step (3) combines the two sums over
\(i\) by distributing each weight \(w_i\) over the reward and next value.
Step (4) interchanges the finite sums over agents and over \((a,s',o')\).  In this step,
the weights are fixed, and the policy, transition, and observation factors
are the same for every reward index \(i\).  Finally, in step (5), the
inner sum for each \(i\) is Equation~\eqref{eq:bellman} in \(\mathcal M\),
so it equals \(V_{i,t}^{\pi;\mathcal M}(s,h^t)\).

Since \(j\) and \((s,h^t)\) were arbitrary, this establishes the identity
at time step \(t\) for every derived agent and valid input.  Starting
from \(t=T\) and repeating this induction step proves it at
\(t=T-1,\ldots,0\).  The policy \(\pi\) was also arbitrary, so the identity
holds for every decentralized policy as stated in the theorem.

We next show that the weight-preserving elements of
\(G_{\mathcal M}^{\mathrm{eval}}\) form a subgroup of
\(G_{\mathcal M}^{\mathrm{eval}}\).  Since
\(G_{\mathcal M}^{\mathrm{eval}}\) is already a group, it suffices to verify
three properties of the set of weight-preserving policy-evaluation
symmetries: (i) the identity transformation belongs to this set;
(ii) composing any two transformations in this set gives another
transformation in the set; and (iii) the inverse of every transformation
in this set also belongs to the set.
The identity preserves \(w\).  If \(g\) and \(k\) preserve the weights, then
\[
w_{\sigma_{gk}(i)}
=w_{\sigma_g(\sigma_k(i))}
=w_{\sigma_k(i)}
=w_i.
\]
Finally, \(w_{\sigma_g(i)}=w_i\) implies the corresponding equality for
\(g^{-1}\) after substituting \(i=\sigma_g^{-1}(k)\).

Now let \(g\in G_{\mathcal M}^{\mathrm{eval}}\) preserve the weights.  For every derived
agent \(j\),
\[
\begin{aligned}
V_{\sigma_g(j),t}^{g\cdot\pi;\mathcal M_w}(\tau_g^t x)
&=\sum_i w_{\sigma_g(i)}
 V_{\sigma_g(i),t}^{g\cdot\pi;\mathcal M}(\tau_g^t x)\\
&=\sum_iw_iV_{i,t}^{\pi;\mathcal M}(x)
=V_{j,t}^{\pi;\mathcal M_w}(x).
\end{aligned}
\]
The first equality applies the weighted-value identity and reindexes the
agent sum by \(k=\sigma_g(i)\).  The second uses weight preservation and
source policy-evaluation equivariance.  Since this holds for every
policy, agent, time step, and valid state--joint-history pair,
\(g\in G_{\mathcal M_w}^{\mathrm{eval}}\), proving
the asserted subgroup inclusion.

At the componentwise level, let \(g\in\AutEnv(\mathcal M)\)
preserve the weights.  By construction, reward aggregation
changes only the reward functions: the transition distribution \(P\) and
observation distribution \(Z\) are unchanged in \(\mathcal M_w\).
Consequently, the preservation identities in
Equations~\eqref{eq:component-P} and~\eqref{eq:component-Z}, which \(g\)
already satisfies in \(\mathcal M\), also hold in \(\mathcal M_w\).
It remains to check preservation of the aggregate reward \(r^w\).
Reindexing the defining reward sum by
\(k=\sigma_g(i)\) gives
\[
\begin{aligned}
r^w(\phi_g^{\mathcal S}(s),g\cdot a,\phi_g^{\mathcal S}(s'))
&=\sum_i w_{\sigma_g(i)}
 r_{\sigma_g(i)}(\phi_g^{\mathcal S}(s),g\cdot a,
                 \phi_g^{\mathcal S}(s'))\\
&=\sum_iw_ir_i(s,a,s')=r^w(s,a,s').
\end{aligned}
\]
Thus it is also a componentwise environment automorphism of \(\mathcal M_w\).

The weight-preserving componentwise environment automorphisms form a subgroup
as well.  The identity preserves the weights.  Closure under composition
follows from the displayed calculation for
\(w_{\sigma_{gk}(i)}\), and the same substitution proves inverse closure.
Proposition~\ref{prop:supp-componentwise-subgroup} establishes that
\(\AutEnv(\mathcal M)\) is a group.  The checks above show that the set of
its weight-preserving elements contains the identity and is closed under
composition and inverses.  These properties establish that this set is a
subgroup of \(\AutEnv(\mathcal M)\).  The preceding componentwise
calculation shows that every element of this subgroup belongs to
\(\AutEnv(\mathcal M_w)\).  Therefore
\[
\left\{g\in\AutEnv(\mathcal M):
w_{\sigma_g(i)}=w_i\ \text{for every }i\right\}
\leq\AutEnv(\mathcal M_w).
\]

\end{proof}

\begin{proposition}[Expected returns under weighted aggregation]
\label{prop:supp-weighted-return}
Under the construction of Theorem~\ref{thm:weighted-aggregation}, the
models \(\mathcal M\) and \(\mathcal M_w\) share the initial distribution
\(\rho_0\).  For every policy \(\pi\in\Pi_{\mathrm{dec}}\) and every
agent \(j\in N\),
\[
J_j^{\mathcal M_w}(\pi)
=\sum_{i\in N}w_iJ_i^{\mathcal M}(\pi).
\]
This identity does not require any symmetry assumption.

Moreover, let \(g\in G_{\mathcal M}^{\mathrm{eval}}\) satisfy
\[
\begin{aligned}
w_{\sigma_g(i)}&=w_i
&&\text{for every }i\in N,\\
\rho_0(\phi_g^{\mathcal S}(s))&=\rho_0(s)
&&\text{for every }s\in\mathcal S.
\end{aligned}
\]
Then, for every policy \(\pi\in\Pi_{\mathrm{dec}}\) and agent \(j\in N\),
\[
J_{\sigma_g(j)}^{\mathcal M_w}(g\cdot\pi)
=J_j^{\mathcal M_w}(\pi).
\]
\end{proposition}

\begin{proof}
Fix any policy \(\pi\in\Pi_{\mathrm{dec}}\) and agent \(j\in N\).
The expected-return definition in Equation~\eqref{eq:expected-return}
and the weighted-value identity in Theorem~\ref{thm:weighted-aggregation}
give
\[
\begin{aligned}
J_j^{\mathcal M_w}(\pi)
&=\sum_{s\in\mathcal S}\rho_0(s)
  V_{j,0}^{\pi;\mathcal M_w}(s,()^N)\\
&=\sum_{s\in\mathcal S}\rho_0(s)\sum_{i\in N}w_i
  V_{i,0}^{\pi;\mathcal M}(s,()^N)\\
&=\sum_{i\in N}w_i\sum_{s\in\mathcal S}\rho_0(s)
  V_{i,0}^{\pi;\mathcal M}(s,()^N)\\
&=\sum_{i\in N}w_iJ_i^{\mathcal M}(\pi).
\end{aligned}
\]
The second equality applies the theorem at the initial time step to each
state and the empty joint history.  The third interchanges the finite
sums over states and agents.  The first and last equalities use the
expected-return definition with the same initial distribution \(\rho_0\).

For the second claim, weight preservation and
Theorem~\ref{thm:weighted-aggregation} imply that
\(g\in G_{\mathcal M_w}^{\mathrm{eval}}\).  By assumption, \(g\) also
preserves \(\rho_0\), which is the initial distribution of
\(\mathcal M_w\) as well as \(\mathcal M\).  Thus
Proposition~\ref{prop:supp-policy-evaluation-subgroup}, applied to
\(\mathcal M_w\), gives
\[
J_{\sigma_g(j)}^{\mathcal M_w}(g\cdot\pi)
=J_j^{\mathcal M_w}(\pi)
\]
for every policy \(\pi\) and agent \(j\), proving the second claim.
\end{proof}

\begin{proposition}[Aggregation can split or merge structural agent types]
\label{prop:supp-aggregation-order}
Under the construction of Theorem~\ref{thm:weighted-aggregation}, each of the
following outcomes can occur for a one-stage, two-agent source model
\(\mathcal M\) and its derived common-reward model \(\mathcal M_w\):
\begin{enumerate}
\item With unequal weights \(w_1\ne w_2\), one structural agent type in
\(\mathcal M\) splits into two distinct structural agent types in
\(\mathcal M_w\).
\item With equal weights \(w_1=w_2\), two distinct structural agent types in
\(\mathcal M\) merge into one structural agent type in \(\mathcal M_w\).
\item With unequal weights \(w_1\ne w_2\), two distinct structural agent
types in \(\mathcal M\) merge into one structural agent type in
\(\mathcal M_w\).
\end{enumerate}
Consequently, neither direction of refinement between
\(\mathcal P(G_{\mathcal M}^{\mathrm{eval}})\) and
\(\mathcal P(G_{\mathcal M_w}^{\mathrm{eval}})\) holds for all source models
\(\mathcal M\) and weight vectors \(w\).
\end{proposition}

\begin{proof}
We construct one example for each of the three claims. All three use one-stage,
two-agent models with singleton
state and observation spaces, binary action sets, deterministic
self-transition, and an admissible group containing only the identity and a
transformation \(g\) whose agent permutation swaps the two agents and whose
local action-label maps are identities.

\noindent\textbf{Case 1: Splitting a Structural Agent Type (Unequal Weights).}
Take
\[
r_1(a_1,a_2)=a_1,\qquad r_2(a_1,a_2)=a_2.
\]
\(g\) is a componentwise environment automorphism of \(\mathcal M\), so its
orbit partition is \(\{\{1,2\}\}\).  With \(w=(3/4,1/4)\), the aggregate
common reward is
\[
\begin{aligned}
r^w(a_1,a_2)
&=\frac34r_1(a_1,a_2)+\frac14r_2(a_1,a_2)\\
&=\frac34a_1+\frac14a_2.
\end{aligned}
\]
Consequently,
\[
r^w(1,0)=\frac34,\qquad r^w(0,1)=\frac14.
\]
For the deterministic policy \(\pi\) selecting \((1,0)\), the transformed
policy \(g\cdot\pi\) selects \((0,1)\).  Hence,
\[
V_{2,0}^{g\cdot\pi;\mathcal M_w}=\frac14
\neq\frac34=V_{1,0}^{\pi;\mathcal M_w}.
\]
Policy-evaluation symmetry requires this identity for every policy, so this
single counterexample is enough to exclude \(g\).
The only nonidentity admissible transformation \(g\) therefore fails the defining
evaluation identity in the derived model, whose partition is
\(\{\{1\},\{2\}\}\).  Aggregation has split a structural agent type of
\(\mathcal M\).

\noindent\textbf{Case 2: Merging Structural Agent Types (Equal Weights).}

\noindent Take
\[
r_1(a_1,a_2)=a_1,\qquad r_2(a_1,a_2)=a_2+1.
\]
Transforming the deterministic policy that selects \((0,0)\) by \(g\) leaves it
unchanged but gives
\(V_{1,0}^{\pi;\mathcal M}=0\) and
\(V_{2,0}^{g\cdot\pi;\mathcal M}=1\).  Evaluation equivariance would require
these values to be equal, so \(g\) is not a policy-evaluation symmetry of
\(\mathcal M\).  Because the admissible group contains only the identity
and \(g\), the partition of \(\mathcal M\) is \(\{\{1\},\{2\}\}\).  With
\(w=(1,1)\), however, the aggregate common reward is
\[
\begin{aligned}
r^w(a_1,a_2)
&=r_1(a_1,a_2)+r_2(a_1,a_2)\\
&=a_1+(a_2+1)\\
&=a_1+a_2+1,
\end{aligned}
\]
which is invariant under \(g\).  The singleton dynamics and
observations are invariant as well, so \(g\) is a componentwise environment
automorphism, hence a policy-evaluation symmetry, in \(\mathcal M_w\).  Its partition is
\(\{\{1,2\}\}\).  Aggregation has merged the two source structural agent types.

\noindent\textbf{Case 3: Merging Structural Agent Types (Unequal Weights).}
Take
\[
r_1(a_1,a_2)=a_1,
\qquad
r_2(a_1,a_2)=\frac12a_2.
\]
For the deterministic policy \(\pi\) selecting \((1,0)\), the transformed
policy \(g\cdot\pi\) selects \((0,1)\), and
\[
V_{2,0}^{g\cdot\pi;\mathcal M}=\frac12
\neq 1=V_{1,0}^{\pi;\mathcal M}.
\]
Thus \(g\notin G_{\mathcal M}^{\mathrm{eval}}\), and the source partition is
\(\{\{1\},\{2\}\}\).  With \(w=(1/3,2/3)\), the aggregate common reward is
\[
r^w(a_1,a_2)
=\frac13r_1(a_1,a_2)+\frac23r_2(a_1,a_2)
=\frac13(a_1+a_2),
\]
which is invariant under \(g\).  The singleton dynamics and observations are
also invariant, so \(g\) is a componentwise environment automorphism and
hence a policy-evaluation symmetry of \(\mathcal M_w\).  Its partition is
\(\{\{1,2\}\}\).  Therefore, aggregation can merge distinct source
structural agent types even when their weights differ.

Cases~1--3 establish the three claimed outcomes. Case~1 and either Case~2 or
Case~3 establish the two opposite failures of refinement, proving the final
claim.
\end{proof}

\section{Detailed Comparisons to Prior Symmetry Frameworks}
\label{app:comparisons}

Sections~\ref{sec:policy-evaluation}--\ref{sec:cooperative} define agent
heterogeneity through the agent orbits of
\(G_{\mathcal M}^{\mathrm{eval}}\).  The comparisons below do not introduce
additional structural agent types.  They instead state how two prior symmetry
frameworks relate to the preservation conditions used in this paper.

\subsection{Counting and Isomorphic Symmetry}

\citet{BraunGL022} formulate counting symmetry and isomorphic symmetry for
partitioned Dec-POMDPs.  Counting symmetry makes the transition,
common-reward, and observation functions invariant to within-partition
permutations of action and observation assignments.  Equivalently, those
functions depend on the assignments only through their per-partition
histograms.  Isomorphic symmetry additionally requires these functions to
factor into repeated identical templates and therefore implies counting
symmetry after grounding, whereas the converse need not hold.  Thus, the two
notions constrain the Dec-POMDP model functions and, for isomorphic symmetry,
their factorization.  By contrast, applying our more general POSG definition
to the resulting grounded Dec-POMDP gives a condition on its complete
policy-evaluation operator: it requires
\(\mathcal E_{\mathcal M}(g\cdot\pi)
=g\cdot\mathcal E_{\mathcal M}(\pi)\) for every decentralized policy.

\paragraph{Lifted-model implication.}
Within Braun et al.'s framework, the implication relates two properties of a
lifted model:
\[
\begin{gathered}
\text{lifted isomorphic factorization}\\[-2pt]
\Downarrow\\[-2pt]
\text{counting symmetry after grounding}.
\end{gathered}
\]
Here, grounding instantiates the lifted factor templates for the individual
agents.  Isomorphic factorization creates identical copies of the same factor
structure for all agents in each part of the lifted partition.  Permuting
agents within a part therefore only reorders identical factors, so the
grounded transition, common-reward, and observation functions can be
represented using within-partition counts.  This is counting symmetry.  The
arrow is one-way: counting symmetry can hold even when the grounded model does
not arise from the same repeated isomorphic factorization.

\paragraph{Consequence for admissible transformations.}
To translate counting symmetry into our language, let
\(\mathcal I=\{I_1,\ldots,I_K\}\) be the lifted agent partition and set
\[
H_{\mathcal I}=\prod_{k=1}^{K}\operatorname{Sym}(I_k),
\qquad
\iota:H_{\mathcal I}\hookrightarrow\mathcal R_{\mathrm{adm}}(\mathcal M),
\qquad
\widehat H_{\mathcal I}:=\iota(H_{\mathcal I}).
\]
The group \(H_{\mathcal I}\) contains exactly the agent permutations
\(\sigma\) that permute agents only within each part \(I_k\).  The map
\(\iota\) assigns every \(\sigma\in H_{\mathcal I}\) a complete admissible
transformation
\(\iota(\sigma)\in\mathcal R_{\mathrm{adm}}(\mathcal M)\), whose agent
permutation is \(\sigma\).  We assume that \(\iota\) is an injective
homomorphism.  Thus distinct within-partition permutations produce distinct
admissible transformations, and composition is preserved:
\(\iota(\sigma_1\sigma_2)=\iota(\sigma_1)\iota(\sigma_2)\).  Consequently,
\(\widehat H_{\mathcal I}=\{\iota(\sigma):\sigma\in H_{\mathcal I}\}\) is a
subgroup of \(\mathcal R_{\mathrm{adm}}(\mathcal M)\) that is isomorphic to
\(H_{\mathcal I}\).  This embedding alone supplies only admissible
transformations.  To make them componentwise environment automorphisms, assume in
addition that every \(\iota(\sigma)\) has the identity state map and
bijectively reindexes every grounded transition, observation, and additive
common-reward factor into an equal instance of the same factor template, with
no asymmetric extra factor.  These assumptions give
the componentwise preservation equations in
Equations~\eqref{eq:component-P}--\eqref{eq:component-r}.  Therefore,
\[
\widehat H_{\mathcal I}
\leq\AutEnv(\mathcal M)
\leq\Geval,
\qquad
\mathcal I
=\mathcal P(\widehat H_{\mathcal I})
\preceq\Peval(\mathcal M).
\]
The first inclusion follows from the stated componentwise preservation
assumptions, and the second follows from
Theorem~\ref{thm:componentwise-certificate}.  Because the agent component of
\(\iota(\sigma)\) is \(\sigma\), the orbits of
\(\widehat H_{\mathcal I}\) are exactly \(I_1,\ldots,I_K\).  Hence, each
lifted partition part is contained in one structural agent type because
\(\widehat H_{\mathcal I}\leq G_{\mathcal M}^{\mathrm{eval}}\).  The
containment may be strict because \(G_{\mathcal M}^{\mathrm{eval}}\) may
contain additional symmetries that connect agents from different lifted
partition parts.  Counting symmetry is thus a
sufficient condition for this particular subgroup of componentwise
environment automorphisms under the stated grounding assumptions.  It is not
a separate structural-type definition and does not claim to recover every
componentwise environment automorphism.  Since each lifted transformation has
the identity state map, it automatically preserves \(\rho_0\), without any
additional restriction on the initial distribution.  Because the initial
joint history is \(()^N\), each
initial state--joint-history pair \((s_0,()^N)\) in
Equation~\eqref{eq:expected-return} is preserved as well.  These facts are needed when
deriving identities for \(J_i(\pi)\), not for the subgroup relation
\(\widehat H_{\mathcal I}\leq G_{\mathcal M}^{\mathrm{eval}}\).

Counting symmetry does not imply the repeated additive factorization
considered here for a partition with one part, even when both notions use the
same lifted partition.  Consider
a one-stage, two-agent Dec-POMDP with state space
\(\mathcal S=\{s_\star\}\), binary action spaces
\(\mathcal A_1=\mathcal A_2=\{0,1\}\), singleton observation spaces,
deterministic self-transition, and common coordination reward
\[
r^{\mathrm{com}}(s_\star,(a_1,a_2),s_\star)
=\mathbf 1\{a_1=a_2\}.
\]
For the lifted partition with one part, \(\mathcal I=\{\{1,2\}\}\), both
agents belong to the same partition part.  The reward
equals \(1\) for the action histograms \((2,0)\) and \((0,2)\), and \(0\)
for \((1,1)\).  It therefore has counting symmetry.  The admissible
transformation whose agent permutation swaps the agents is also a componentwise
environment automorphism and hence a policy-evaluation symmetry.  However,
because the model has only the fixed state \(s_\star\), abbreviate
\(r^{\mathrm{com}}(s_\star,(a_1,a_2),s_\star)\) by
\(r^{\mathrm{com}}(a_1,a_2)\).  For this same partition with one part, the
repeated additive factorization considered here uses the same local factor template
\(u\) for both agents:
\(r^{\mathrm{com}}(a_1,a_2)=c+u(a_1)+u(a_2)\).  The shared function \(u\)
is the repeated template associated with the only part \(\{1,2\}\).  This form
would require
\[
r^{\mathrm{com}}(0,0)+r^{\mathrm{com}}(1,1)
=r^{\mathrm{com}}(0,1)+r^{\mathrm{com}}(1,0),
\]
because both sides would equal \(2c+2u(0)+2u(1)\).  Although the coordination
reward satisfies counting symmetry, the left-hand side is \(1+1=2\), whereas
the right-hand side is \(0+0=0\).  Hence, no choices of \(c\) and \(u\) give
this reward in the stated repeated additive form.  Thus counting symmetry
does not imply this repeated additive factorization for a partition with one
part. More
generally, policy-evaluation symmetry does not by itself provide the
repeated lifted factor structure used to obtain the compact representation
and tractability guarantees for lifted isomorphic models.
Section~\ref{app:lifted-isomorphism} gives the factor-reindexing proof and
the proof that this example lacks the stated factorization for a partition
with one part.

\subsubsection{Proofs of the lifted-model comparison}
\label{app:lifted-isomorphism}
The first proposition makes the grounding assumptions explicit; the second
establishes the same-partition non-converse.

\begin{proposition}[Lifted symmetry as a sufficient condition]
\label{prop:supp-lifted-certificate}
Let \(\mathcal I=\{I_1,\ldots,I_K\}\) be an agent partition and let
\[
H_{\mathcal I}=\prod_{k=1}^K\operatorname{Sym}(I_k).
\]
Suppose a finite Dec-POMDP grounding admits an injective homomorphism
\[
\iota:H_{\mathcal I}\longrightarrow\mathcal R_{\mathrm{adm}}(\mathcal M)
\]
whose image consists of transformations with the corresponding agent permutation, identity
state map, and corresponding action and observation coordinate maps.  Suppose
also that every transformation in the image bijectively reindexes all grounded
transition, observation, and additive common-reward factor instances into
instances of the same templates, with no asymmetric extra factors.  Writing
\(\widehat H_{\mathcal I}=\iota(H_{\mathcal I})\), we then have
\[
\widehat H_{\mathcal I}
\leq\AutEnv(\mathcal M)
\leq G_{\mathcal M}^{\mathrm{eval}},
\qquad
\mathcal I=\mathcal P(\widehat H_{\mathcal I})
\preceq\mathcal P(G_{\mathcal M}^{\mathrm{eval}}).
\]
Because the state map is the identity and the initial history is empty, the
same transformations preserve the initial distribution.
\end{proposition}

\begin{proof}
Fix a within-partition permutation \(\sigma\in H_{\mathcal I}\) and its
complete transformation \(\iota(\sigma)\).  By assumption, it merely
reorders the finitely many grounded factors within each factor product
defining a transition or observation probability and within the factor sum
defining the common reward.  Every factor is matched with an equal instance
of the same template.  Reordering a finite product or sum does not change its
value.  Thus
\(\iota(\sigma)\) preserves the grounded transition distribution,
observation distribution, and reward componentwise.  Because this argument
holds for every \(\sigma\),
\(\widehat H_{\mathcal I}\leq\AutEnv(\mathcal M)\).  Theorem~\ref{thm:componentwise-certificate}
then gives \(\AutEnv(\mathcal M)\leq G_{\mathcal M}^{\mathrm{eval}}\).

We next show that the agent orbits of \(\widehat H_{\mathcal I}\) are exactly
the parts of \(\mathcal I\). Fix a part \(I_k\) and an agent \(i\in I_k\).
Every \(g\in\widehat H_{\mathcal I}\) has the form \(g=\iota(\sigma)\) for
some \(\sigma\in H_{\mathcal I}\), with agent permutation
\(\sigma_g=\sigma\). Since \(\sigma\) permutes agents only within each
part, \(\sigma_g(i)\in I_k\). Therefore,
\[
\{\sigma_g(i):g\in\widehat H_{\mathcal I}\}\subseteq I_k.
\]
Conversely, take any \(j\in I_k\). The permutation that exchanges \(i\)
and \(j\) and fixes all other agents belongs to \(H_{\mathcal I}\), because
it acts only within \(I_k\); if \(j=i\), use the identity permutation.
For this permutation \(\sigma\), its image
\(\iota(\sigma)\in\widehat H_{\mathcal I}\) satisfies
\(\sigma_{\iota(\sigma)}(i)=\sigma(i)=j\). Thus every agent in \(I_k\)
belongs to the orbit of \(i\), proving the reverse containment. Hence,
\[
\{\sigma_g(i):g\in\widehat H_{\mathcal I}\}=I_k.
\]
Since this holds for every \(k\) and every \(i\in I_k\), the orbit partition
is \(\mathcal P(\widehat H_{\mathcal I})=\mathcal I\).

To establish the refinement, use the inclusion
\(\widehat H_{\mathcal I}\leq G_{\mathcal M}^{\mathrm{eval}}\) proved above.
Every transformation used to connect two agents in \(I_k\) therefore also belongs
to \(G_{\mathcal M}^{\mathrm{eval}}\). Consequently, each part \(I_k\) is
contained in one \(G_{\mathcal M}^{\mathrm{eval}}\)-orbit, which is one
structural agent type. By the definition of partition refinement, as in
Proposition~\ref{prop:supp-orbit-facts}, this gives
\[
\mathcal I=\mathcal P(\widehat H_{\mathcal I})
\preceq\mathcal P(G_{\mathcal M}^{\mathrm{eval}}).
\]

Finally, the identity state map fixes the initial
state coordinate and every transformation fixes the empty initial history.
Hence each transformation in the image preserves the initial distribution as claimed.
\end{proof}

\begin{proposition}[Same-partition non-converse]
\label{prop:supp-lifted-nonconverse}
In the one-stage, two-agent common-reward model with binary actions and
\[
R(a_1,a_2)=\mathbf 1\{a_1=a_2\},
\]
the candidate partition with one part, \(\{\{1,2\}\}\), has counting
symmetry, and
the transformation whose agent permutation swaps the two agents is a
componentwise environment automorphism and therefore a policy-evaluation
symmetry.  However, there are no
constant \(c\) and repeated local contribution \(u\) such that
\[
R(a_1,a_2)=c+u(a_1)+u(a_2).
\]
\end{proposition}

\begin{proof}
The reward is determined by the action histogram for the only partition
part: it is \(1\) for \((2,0)\) and \((0,2)\), and \(0\) for \((1,1)\).  Together with the singleton
state and observation spaces and deterministic self-transition, this proves
counting symmetry.  The reward is therefore unchanged by exchanging the two
action coordinates, while every other component is also unchanged.  The
corresponding transformation, whose agent permutation swaps the two agents,
is a componentwise environment automorphism, and
Theorem~\ref{thm:componentwise-certificate} shows that it is a
policy-evaluation symmetry.

If the repeated additive form held, then
\[
\begin{aligned}
R(0,0)+R(1,1)
&=[c+2u(0)]+[c+2u(1)]\\
&=2[c+u(0)+u(1)]
=R(0,1)+R(1,0).
\end{aligned}
\]
For the coordination reward, the left side is \(2\) and the right side is
\(0\), a contradiction.  This rules out the stated repeated additive reward
factorization for a partition with one part.  It does not rule out a partition
into singleton parts, which separates the agents in the lifted representation.
\end{proof}

\subsection{Expected Return Symmetry}

An expected return symmetry (ERS) is defined through self-play-optimal
Boltzmann policies at a fixed temperature \(\alpha>0\)
\citep{muglich2025expected}.  We compare the two notions on their shared
subclass of finite-horizon Dec-POMDPs: initial histories are empty, local
action spaces are fixed, all agents have a common reward function,
and the observation distribution factorizes as
\(Z(o'\mid s',a)=\prod_i U_i(o_i'\mid s',a)\).  Let \(\Psi\) denote the
ambient transformation group used by ERS.  We identify each of its elements with
the type-compatible transformation
\[
g=\bigl(\operatorname{id}_N,\phi_g^{\mathcal S},
(\phi_{g,i}^{\mathcal A})_{i\in N},
(\phi_{g,i}^{\mathcal O})_{i\in N}\bigr),
\]
where \(\phi_g^{\mathcal S}:\mathcal S\to\mathcal S\),
\(\phi_{g,i}^{\mathcal A}:\mathcal A_i\to\mathcal A_i\), and
\(\phi_{g,i}^{\mathcal O}:\mathcal O_i\to\mathcal O_i\) are bijections.
The group \(\Psi\) contains all tuples of this form.  Thus, \(\Psi\) fixes
the agent indices.  Under this identification, \(\Psi\)
and \(\mathcal R_{\mathrm{adm}}(\mathcal M)\) are groups of type-compatible transformations
on the same model, so their intersection is well defined.  They act on
decentralized policies by the policy-transformation rule in
Equation~\eqref{eq:policy-action}.  Let \(\Pi^\alpha(\mathcal M)\) be the
policies satisfying the Boltzmann condition at every local history, stated
explicitly in Section~\ref{app:expected-return}, and let
\[
\Pi_*^\alpha(\mathcal M)
=\operatorname*{arg\,max}_{\pi\in\Pi^\alpha(\mathcal M)}
J^{\mathcal M}(\pi),
\]
where \(J^{\mathcal M}(\pi)\) is the common return under the initial
distribution.  We compare the two notions only on
\(\Psi\cap\mathcal R_{\mathrm{adm}}(\mathcal M)\), whose elements are
admissible in both formalisms.  Throughout this subsection, the restricted
ERS group is expressed, as follows:
\[
\Phi_{\mathrm{ER}}^\alpha(\mathcal M)
\coloneqq
\bigl\{g\in\Psi\cap\mathcal R_{\mathrm{adm}}(\mathcal M):\;
\{g\cdot\pi:\pi\in\Pi_*^\alpha(\mathcal M)\}
=\Pi_*^\alpha(\mathcal M)\bigr\}.
\]
Here, \(\{g\cdot\pi:\pi\in\Pi_*^\alpha(\mathcal M)\}\) is the image of
the entire self-play-optimal Boltzmann policy set under \(g\).  The equality
therefore requires \(g\) to map this set onto itself, although it may map one
member to a different member.  Because the state, action, and observation
spaces are finite, the common transformation domain is a finite group.  The
set equality is consequently equivalent to the pointwise condition used by
\citet{muglich2025expected}, namely
\(g\cdot\pi\in\Pi_*^\alpha(\mathcal M)\) for every
\(\pi\in\Pi_*^\alpha(\mathcal M)\).
Thus, ERS concerns a selected optimal Boltzmann policy set and its local
action-value conditions.  By contrast, policy-evaluation symmetry preserves
complete policy evaluation under a type-compatible transformation for every decentralized
policy and may have a nonidentity agent permutation.

\begin{proposition}[Class-level two-way non-implication]
\label{prop:ers-noncontainment}
For every \(\alpha>0\), there are finite Dec-POMDPs
\(\mathcal M_a,\mathcal M_b\) and transformations \(g_a,g_b\) admitted by
both formalisms such that
\[
g_a\in
\Phi_{\mathrm{ER}}^\alpha(\mathcal M_a)
\setminus G_{\mathcal M_a}^{\mathrm{eval},\rho_0},
\qquad
g_b\in
G_{\mathcal M_b}^{\mathrm{eval},\rho_0}
\setminus\Phi_{\mathrm{ER}}^\alpha(\mathcal M_b).
\]
Hence neither condition implies the other across the common model class.
\end{proposition}

\paragraph{Proof sketch.}
The proposition gives two counterexamples over the common model class.  It
does not claim that both set differences are nonempty for every fixed
Dec-POMDP.  In \(\mathcal M_a\), the uniform policy is the unique
self-play-optimal Boltzmann policy.  A within-agent action relabeling fixes
this singleton set but changes the return of a deterministic policy outside
\(\Pi^\alpha\).  It is therefore an ERS transformation but not a
policy-evaluation symmetry.  In \(\mathcal M_b\), a within-agent observation
relabeling satisfies the value-transformation identity in
Definition~\ref{def:policy-evaluation} for every decentralized policy after
transforming the policy and local histories.
However, it exchanges observations with different conditional local
action-value gaps, so it maps the unique Boltzmann policy to a policy that
violates Boltzmann consistency.  Both transformations have the identity state
map and therefore preserve the corresponding initial distribution.
Section~\ref{app:expected-return} gives both constructions.

To state the exact additional condition, let
\(Q_i^\pi(h_i^t,a_i)\) be the conditional expected return-to-go when agent
\(i\)'s current action is fixed to \(a_i\), while the other agents' current
actions and all later actions follow \(\pi\).  Consider
\(g\in G_{\mathcal M}^{\mathrm{eval},\rho_0}\cap\Psi\).  Assume that
\(\Phi_{g,i}^{\mathcal H,t}(h_i^t)\) lies in the paired local-history domain,
\(\phi_{g,i}^{\mathcal A}\) bijects the corresponding legal-action sets, and
both local action-value vectors are defined.
Proposition~\ref{prop:supp-local-q-bridge} shows that
\(g\in\Phi_{\mathrm{ER}}^\alpha(\mathcal M)\) if and only if, for
every \(\pi\in\Pi_*^\alpha(\mathcal M)\), agent \(i\), time step \(t\), and local
history \(h_i^t\), there is a constant \(c_{g,\pi,i,h_i^t}\), independent of
\(a_i\), such that:
\[
Q_i^{g\cdot\pi}
\left(
\Phi_{g,i}^{\mathcal H,t}(h_i^t),
\phi_{g,i}^{\mathcal A}(a_i)
\right)
=Q_i^\pi(h_i^t,a_i)+c_{g,\pi,i,h_i^t}
\]
for every legal \(a_i\).  The constant is exactly the freedom that leaves a
softmax distribution unchanged.  If, more strongly, such a \(g\) is also a
componentwise environment automorphism, the constant is zero at histories
having positive probability under the paired policies.  At an off-path
history, meaning a history with zero probability under the relevant policy,
the same conclusion requires one globally fixed belief and local-\(Q\)
convention that is equivariant under the transformations being compared.

Because ERS as defined by \citet{muglich2025expected} fixes agent indices, it
does not relate distinct agents and therefore cannot replace the structural
agent types defined in this paper.  Those types remain the agent orbits of the
policy-evaluation symmetry group \(G_{\mathcal M}^{\mathrm{eval}}\), as in
Definition~\ref{def:structural-type}.

\subsubsection{Boltzmann condition and comparison proofs}
\label{app:expected-return}

For the proofs below, fix a temperature \(\alpha>0\) and use the common
admissible transformation domain defined above. We abbreviate
\(\Pi^\alpha(\mathcal M)\), \(\Pi_*^\alpha(\mathcal M)\), and
\(J^{\mathcal M}(\pi)\) by \(\Pi^\alpha\), \(\Pi_*^\alpha\), and
\(J(\pi)\), respectively. Recall that \(\Pi^\alpha\) contains the policies
satisfying the Boltzmann condition
\[
\pi_i^t(a_i\mid h_i^t)
=
\frac{\exp(Q_i^\pi(h_i^t,a_i)/\alpha)}
{\sum_{b_i\in\mathcal A_i}
 \exp(Q_i^\pi(h_i^t,b_i)/\alpha)}.
\]
The local action value
\(
Q_i^\pi(h_i^t,a_i)
\)
is the conditional expected return-to-go after local history \(h_i^t\) when
agent \(i\)'s current action is fixed to \(a_i\), while the other current
actions and all later actions follow \(\pi\).  At a zero-probability history
it uses the single declared off-path belief and local-\(Q\) convention
discussed below.

\begin{proof}[Proof of Proposition~\ref{prop:ers-noncontainment}]
\emph{ERS without policy-evaluation symmetry.}
Choose \(0<\varepsilon<2\alpha\).  Let \(\mathcal M_a\) be a one-stage,
two-agent Dec-POMDP with one state, singleton observations, binary actions,
deterministic dynamics, and common reward
\(
R(a)=\varepsilon\,\mathbf 1\{a_1=a_2\}.
\)
If agent \(i\) selects action \(1\) with probability \(p_i\), then against
an agent selecting action \(1\) with probability \(q\),
\[
Q_i^\pi((),1)=\varepsilon q,
\qquad
Q_i^\pi((),0)=\varepsilon(1-q).
\]
Boltzmann self-consistency is therefore
\[
p_1=F(p_2),\qquad p_2=F(p_1),\qquad
F(q)=\frac{1}{1+\exp[-\varepsilon(2q-1)/\alpha]}.
\]
Since
\[
|F'(q)|\leq\frac{\varepsilon}{2\alpha}<1,
\]
the joint response map
\((p_1,p_2)\mapsto(F(p_2),F(p_1))\) is a contraction in the sup norm.
Moreover, \(F(1/2)=1/2\), so \((1/2,1/2)\) is a fixed point.  To see directly
that it is the only one, let \(p\) be any other fixed point and let
\(L=\varepsilon/(2\alpha)<1\).  The contraction bound gives
\[
\lVert p-(1/2,1/2)\rVert_\infty
\leq L\lVert p-(1/2,1/2)\rVert_\infty,
\]
which forces the norm to be zero.  Therefore
\[
\Pi^\alpha=\Pi_*^\alpha=\{\pi^{\mathrm u}\},
\]
where \(\pi^{\mathrm u}\) is the uniform product policy.

Let \(g_a\) swap the two action labels of agent \(1\) and fix all other
coordinates.  Declare \(\{e,g_a\}\) admissible.  This within-agent action
relabeling lies in \(\Psi\) and fixes \(\pi^{\mathrm u}\), hence
\[
g_a\in\Phi_{\mathrm{ER}}^\alpha(\mathcal M_a).
\]
For the deterministic policy \(\pi^{00}\), however,
\[
J^{\mathcal M_a}(\pi^{00})=\varepsilon,
\qquad
J^{\mathcal M_a}(g_a\cdot\pi^{00})=0.
\]
The transformation fixes the agent indices and the unique initial pair, so
return equivariance would require these two returns to be equal.  Their
inequality therefore violates policy-evaluation symmetry, and
\[
g_a\notin G_{\mathcal M_a}^{\mathrm{eval},\rho_0}.
\]

\emph{Policy-evaluation symmetry without ERS.}
Let \(\mathcal M_b\) be a two-stage, two-agent Dec-POMDP.  Use states
\(\{u,z\}\).  Agent \(1\) has actions \(\{0,1\}\) and observations
\(\{c,x,y\}\).  Agent \(2\) has singleton action and observation sets
\(\{\star\}\) and \(\{\bot\}\).  Set \(\gamma=1\), let the initial
distribution be \(\delta_u\), and abbreviate joint action \((a,\star)\) by
\(a\).  For either action, take
\[
P(u\mid u,a)=P(z\mid u,a)=\frac12,
\qquad
P(z\mid z,a)=1.
\]
Agent \(2\) always observes \(\bot\), while agent \(1\)'s
action-independent observation probabilities are
\[
\begin{array}{c|ccc}
 &c&x&y\\ \hline
u&1/5&2/5&2/5\\
z&1/5&1/5&3/5
\end{array}
\]
Give both agents the common successor-state reward
\[
R(s',(a,\star))=2\,\mathbf 1\{a=0,\ s'=u\}.
\]
Let \(g_b\) exchange agent \(1\)'s observation labels \(x\) and \(y\) and
fix every other coordinate.  Declare \(\{e,g_b\}\) admissible.  This
within-agent observation relabeling lies in \(\Psi\).

For \(a^0\in\{0,1\}\), \(o\in\{c,x,y\}\), and either common-reward
coordinate \(k\), write
\[
h^1(a^0,o)=\bigl((a^0,o),(\star,\bot)\bigr),
\qquad
p_{a^0,o}=\pi_1^1(0\mid(a^0,o)).
\]
At \(t=1\),
\[
V_{k,1}^{\pi}\bigl(u,h^1(a^0,o)\bigr)=p_{a^0,o},
\qquad
V_{k,1}^{\pi}\bigl(z,h^1(a^0,o)\bigr)=0.
\]
Indeed, from \(u\), choosing action \(0\) gives reward \(2\) when the next
state is \(u\), which occurs with probability \(1/2\), and choosing action
\(1\) gives zero.  From \(z\), both actions give zero.
The policy transformation exchanges \(p_{a^0,x}\) and \(p_{a^0,y}\).  For
example,
\[
V_{k,1}^{g_b\cdot\pi}
\bigl(u,g_b\cdot h^1(a^0,x)\bigr)
=p_{a^0,x}
=V_{k,1}^{\pi}\bigl(u,h^1(a^0,x)\bigr).
\]
The same calculation applies to histories ending in \(y\), while \(c\) is
fixed and both values from \(z\) are zero.  Hence the defining
value-transformation identity holds at every state--joint-history pair at
\(t=1\).  At the initial time step \(t=0\),
putting
\(q=\pi_1^0(0\mid())\) gives
\[
V_{k,0}^\pi(u,()^N)
=q+\frac12\sum_{a^0\in\{0,1\}}\pi_1^0(a^0\mid())
\left[
\frac15p_{a^0,c}+\frac25p_{a^0,x}+\frac25p_{a^0,y}
\right].
\]
Exchanging \(x\) and \(y\) leaves this expression unchanged.  From state
\(z\), every present and future reward is zero.  Together with the
zero values at \(t=2\), these calculations cover all time steps and valid
state--joint-history pairs.  The transformation \(g_b\) also fixes the state,
so it preserves the initial distribution \(\delta_u\).  Therefore
\[
g_b\in G_{\mathcal M_b}^{\mathrm{eval},\rho_0}.
\]

The local action values used by the Boltzmann condition instead condition on
agent \(1\)'s observation.  For example, after either initial action \(a^0\),
\[
\Pr(u\mid a^0,x)
=\frac{(1/2)(2/5)}{(1/2)(2/5)+(1/2)(1/5)}
=\frac23.
\]
The same Bayes calculation gives
\[
\Pr(u\mid a^0,y)=\frac25,\qquad
\Pr(u\mid a^0,c)=\frac12.
\]
These posterior probabilities are also the action-value gaps at \(t=1\):
conditioned on a belief assigning probability \(p\) to \(u\), action \(0\)
has expected reward \(2p(1/2)=p\), whereas action \(1\) has value zero.
Thus the required Boltzmann probabilities of action \(0\) after \(x\) and
\(y\) are
\[
b_x=\frac{e^{2/(3\alpha)}}{1+e^{2/(3\alpha)}},
\qquad
b_y=\frac{e^{2/(5\alpha)}}{1+e^{2/(5\alpha)}}.
\]
Every initial action and every resulting observation at \(t=1\)
have positive probability under a Boltzmann policy, so the displayed
probabilities uniquely fix every rule at \(t=1\).  The posterior
calculation, and hence each of these fixed rules, is independent of the
initial action.  Since the transition and observation distributions and the
decision rules at \(t=1\) are independent of the initial action, the expected
reward at \(t=1\) is the same after initial actions \(0\) and \(1\).  Their immediate expected
rewards differ by
\(2(1/2)=1\), which uniquely fixes the Boltzmann rule at the initial time step as well.  Hence
\(\Pi^\alpha=\Pi_*^\alpha\) is a singleton.  Since the logistic function is
strictly increasing, \(b_x>b_y\).  At a history ending in \(x\), the
transformed policy uses \(b_y\), although Boltzmann self-consistency requires
\(b_x\).  Therefore
\[
g_b\notin\Phi_{\mathrm{ER}}^\alpha(\mathcal M_b),
\]
which proves the reverse non-implication.
\end{proof}

\begin{proposition}[Local action-value bridge]
\label{prop:supp-local-q-bridge}
Let \(g\) be a transformation such that
\[
g\in G_{\mathcal M}^{\mathrm{eval},\rho_0}\cap\Psi.
\]
Assume that \(g\) preserves the paired local-history domain,
matches the legal-action sets bijectively, and that both local action-value
vectors are defined.  Then \(g\in\Phi_{\mathrm{ER}}^\alpha(\mathcal M)\) if
and only if, for every \(\pi\in\Pi_*^\alpha\), \(i,t,h_i^t\), there is an
action-independent constant \(c_{g,\pi,i,h_i^t}\) such that
\[
Q_i^{g\cdot\pi}
\left(
\Phi_{g,i}^{\mathcal H,t}(h_i^t),
\phi_{g,i}^{\mathcal A}(a_i)
\right)
=Q_i^\pi(h_i^t,a_i)+c_{g,\pi,i,h_i^t}
\]
for every legal \(a_i\).  If, in addition,
\(g\in\AutEnv(\mathcal M)\), then the identity holds with constant zero at
positive-probability paired histories.  The same conclusion
at off-path histories requires a globally fixed belief and local-\(Q\)
convention that is equivariant under the relevant group.
\end{proposition}

\begin{proof}
Because \(g\) is an initial-distribution-preserving policy-evaluation symmetry
and \(g\in\Psi\) fixes agent indices,
\[
J(g\cdot\pi)=J(\pi)
\]
for every policy.  Thus, for \(\pi\in\Pi_*^\alpha\), the transformed policy
is again optimal within \(\Pi^\alpha\) exactly when it remains in
\(\Pi^\alpha\).

The policy transformation preserves the action-probability vector after
reindexing its action coordinates.
Two finite Boltzmann vectors agree exactly when their action values differ by
one action-independent constant.  Write the corresponding action-value
vectors as \(q\) and \(q'\), indexed by corresponding legal actions.
For a singleton action set the claim is
immediate.  Otherwise, equality of the probability ratios for any two actions
\(a,b\) gives
\[
e^{(q'_a-q'_b)/\alpha}=e^{(q_a-q_b)/\alpha},
\]
and hence \(q'_a-q_a=q'_b-q_b\).  Thus the difference is the same for every
action.  Conversely, if \(q'_a=q_a+c\) for every action \(a\), then
\[
\begin{aligned}
\frac{e^{q'_a/\alpha}}{\sum_b e^{q'_b/\alpha}}
&=\frac{e^{c/\alpha}e^{q_a/\alpha}}
        {e^{c/\alpha}\sum_b e^{q_b/\alpha}}\\
&=\frac{e^{q_a/\alpha}}{\sum_b e^{q_b/\alpha}}.
\end{aligned}
\]
The common factor therefore cancels, giving the same action probabilities.

To prove necessity, suppose that
\(g\in\Phi_{\mathrm{ER}}^\alpha(\mathcal M)\).  For every
\(\pi\in\Pi_*^\alpha\), its image \(g\cdot\pi\) also belongs to
\(\Pi_*^\alpha\), so both policies satisfy the Boltzmann condition.
Their action probabilities agree after reindexing the action coordinates.  Applying the
probability-ratio argument at each paired local history gives the stated
action-independent difference between their local action values.

To prove sufficiency, assume the stated local action-value identity for
every \(\pi\in\Pi_*^\alpha\) and every paired local history.  Fix such a
policy \(\pi\).  At each paired history, the constant-shift calculation
shows that its reindexed action probabilities are the Boltzmann probabilities
determined by \(Q_i^{g\cdot\pi}\).  These are precisely the probabilities
assigned by \(g\cdot\pi\), so the transformed policy satisfies the
Boltzmann condition at every local history.  Thus
\(g\cdot\pi\in\Pi^\alpha\).  Since
\(J(g\cdot\pi)=J(\pi)\), it attains the same maximum return within
\(\Pi^\alpha\) as \(\pi\), and hence belongs to \(\Pi_*^\alpha\).
Since \(\pi\) was arbitrary, this proves
\[
\{g\cdot\pi:\pi\in\Pi_*^\alpha\}\subseteq\Pi_*^\alpha.
\]

To prove the reverse inclusion, we must show that every policy in
\(\Pi_*^\alpha\) is the image of another policy in that set.  Because
\(g\) belongs to the finite group \(\Psi\), there is an integer \(m\geq1\)
such that \(g^m=e\), where \(e\) is the identity transformation.  Take any
\(\pi\in\Pi_*^\alpha\) and define
\[
\widetilde\pi=g^{m-1}\cdot\pi.
\]
The inclusion already proved means that each application of \(g\) keeps
a policy in \(\Pi_*^\alpha\).  Applying it \(m-1\) times therefore gives
\(\widetilde\pi\in\Pi_*^\alpha\); when \(m=1\), this simply means
\(\widetilde\pi=\pi\).  Moreover, the policy action law gives
\[
g\cdot\widetilde\pi
=g\cdot(g^{m-1}\cdot\pi)
=g^m\cdot\pi
=\pi.
\]
Thus every \(\pi\in\Pi_*^\alpha\) has a preimage in \(\Pi_*^\alpha\),
which establishes the reverse inclusion.  Combining the two inclusions gives
\[
\{g\cdot\pi:\pi\in\Pi_*^\alpha\}=\Pi_*^\alpha.
\]
This is exactly the set-preservation requirement in the definition of
\(\Phi_{\mathrm{ER}}^\alpha(\mathcal M)\), completing the equivalence proof.

For the zero-constant statement, pair each trajectory under \(\pi\) with its
transformed counterpart under \(g\cdot\pi\).  Componentwise preservation, the policy
transformation, and preservation of the initial distribution give the paired
trajectories the same probability and reward.  Conditioning at a
positive-probability paired history preserves this correspondence.  It also
persists when the current local action is fixed and all other current and
later actions follow the respective policies.  The two conditional expected returns-to-go are
therefore equall, such that:
\[
Q_i^{g\cdot\pi}
\left(
\Phi_{g,i}^{\mathcal H,t}(h_i^t),
\phi_{g,i}^{\mathcal A}(a_i)
\right)
=Q_i^\pi(h_i^t,a_i).
\]
Positive reachability makes the Bayesian conditional law unambiguous.  At a
zero-probability history, this reasoning applies only when the single
off-path belief and local-\(Q\) convention is itself equivariant.  A convention
may assign well-defined beliefs and local action values at every
zero-probability history without making these assignments consistent with
the transformation \(g\) (see the below explanation for details).  Therefore, well-definedness alone does not establish
the local action-value equality at those histories.
\end{proof}

\paragraph{Clarification on Off-Path Histories.}
An off-path history can occur even when every action has positive probability.
For example, consider a two-stage, two-agent Dec-POMDP with states
\(\{L,R\}\), a uniform initial distribution, and transitions that leave the
state unchanged.  Agent \(1\) has actions \(\{0,1\}\) and observations
\(\{x,y\}\), but always observes \(x\) in either state; agent \(2\) has
singleton action and observation sets.  The common reward is \(1\) when
the state and agent \(1\)'s action are \((L,0)\) or \((R,1)\), and \(0\)
otherwise.  At \(t=1\), the history \(h_1^1=(0,y)\) has probability zero
under every policy, because \(y\) is impossible.  To evaluate an action after
this history, agent \(1\) needs a conditional belief about the state.
However, the usual conditional-probability formula would require the
undefined ratio, such that:
\[
\frac{\Pr_\pi(s_1=L,\,a_1^0=0,\,o_1^1=y)}
     {\Pr_\pi(a_1^0=0,\,o_1^1=y)}
=\frac{0}{0}.
\]
Thus, the model and policy do not determine which belief to use at this
history.  To define the local action values and impose the Boltzmann
condition at every local history, we must declare a convention that assigns
a belief at such zero-probability histories.  For example, at \(t=1\),
assigning probability one to \(L\) gives values \(1\) and \(0\) for actions
\(0\) and \(1\), while assigning probability one to \(R\) reverses these values.
These are alternative extensions
to an impossible history, and neither assignment is determined by Bayes' rule.
Let \(g\) exchange
\(L\) with \(R\) and agent \(1\)'s actions \(0\) with \(1\), while fixing
all other coordinates, and take \(\{e,g\}\) as the admissible group.
This transformation preserves the transition, observation, and common-reward
functions and the initial distribution, and maps \((0,y)\) to \((1,y)\).
Suppose a fixed convention assigns probability one to \(L\) at every
zero-probability local history of agent \(1\), under every policy.  Since
\(t=1\) is the final decision time, the resulting local action values satisfy
\[
Q_1^\pi((0,y),0)=1,
\qquad
Q_1^{g\cdot\pi}((1,y),1)=0.
\]
Thus, this well-defined convention does not preserve the local action-value
equality.  An equivariant convention must instead assign probability one to
\(R\) at the transformed history under \(g\cdot\pi\) whenever it assigns
probability one to \(L\) at the source history under \(\pi\).  This issue
concerns the conditional local action values used in the ERS comparison.
The complete values \(V_{i,t}^\pi(s,h^t)\) used to define our symmetry group
are specified directly by Equation~\eqref{eq:bellman}, including at
unreachable state--joint-history pairs, and require no such belief convention.

\end{document}